\documentclass[a4paper,11pt]{article}
\pdfoutput=1 

\usepackage{jcappub} 

\usepackage[T1]{fontenc}
\usepackage{epsfig,latexsym,cancel,amssymb,amsmath}
\usepackage{hyperref}      
\usepackage{graphicx}

\usepackage[caption=false]{subfig}
\usepackage{epstopdf}
\usepackage{color}
\usepackage{bm}
\usepackage[normalem]{ulem}
\usepackage{array}
\usepackage{multirow, multicol}

\renewcommand{\thetable}{$\arabic{table}$}

\DeclareMathOperator{\hyphen}{-}

\newcommand{\gev}{{\rm GeV}}
\newcommand{\kpc}{{\rm kpc}}
\newcolumntype{A}{ >{\centering\arraybackslash} m{3.5cm} }
\newcolumntype{B}{ >{\centering\arraybackslash} m{2.5cm} }
\newcolumntype{C}{ >{\centering\arraybackslash} m{2cm} }
\newcolumntype{D}{ >{\centering\arraybackslash} m{1.5cm} }
\newcolumntype{E}{ >{\centering\arraybackslash} m{2.5cm} }

\def\beq{\begin{equation}}
\def\eeq{\end{equation}}
\def\bea{\begin{eqnarray}}
\def\eea{\end{eqnarray}}

\def\Eq#1{Eq.~(\ref{#1})}

\begin{document}
\hfill{\small MITP/16-075 \, IITB-PHY160916} \\

\title{Late decaying 2-component dark matter scenario as an explanation of the AMS-02 positron excess}

\author[a, b]{Jatan Buch}
\emailAdd{jatan\textunderscore buch@brown.edu}
\affiliation[a]{Department of Physics, Indian Institute of Technology - Bombay, Powai, Mumbai 400076, India}
\affiliation[b]{PRISMA Cluster of Excellence, 55099 Mainz, Germany and Mainz Institute for Theoretical Physics, Johannes Gutenberg-Universit\"{a}t
Mainz, 55099 Mainz, Germany}

\author[a]{Pranjal Ralegankar}
\emailAdd{pranjal6@illinois.edu}

\author[a]{Vikram Rentala}
\emailAdd{rentala@phy.iitb.ac.in}


\abstract{The long standing anomaly in the positron flux as measured by the {\fontfamily{cmss}\selectfont {PAMELA}} and {\fontfamily{cmss}\selectfont {AMS-02}} experiments could potentially be explained by dark matter (DM) annihilations. This scenario typically requires a large ``boost factor'' to be consistent with a thermal relic dark matter candidate produced via freeze-out. However, such an explanation is disfavored by constraints from CMB observations on energy deposition during the epoch of recombination. We discuss a scenario called late-decaying two-component dark matter (LD2DM), where the entire DM consists of two semi-degenerate species. Within this framework, the heavier species is produced as a thermal relic in the early universe and decays to the lighter species over cosmological timescales. Consequently, the lighter species becomes the DM which populates the universe today. We show that annihilation of the lighter DM species with an enhanced cross-section, produced via such a non-thermal mechanism, can explain the observed {\fontfamily{cmss}\selectfont {AMS-02}} positron flux while avoiding CMB constraints. The observed DM relic density can be correctly reproduced as well with simple $s$-wave annihilation cross-sections. We demonstrate that the scenario is safe from CMB constraints on late-time energy depositions during the cosmic ``dark ages''.  Interestingly, structure formation constraints force us to consider small mass splittings between the two dark matter species. We explore possible cosmological and particle physics signatures in a toy model that realizes this scenario.}

\maketitle


\section{Introduction and motivation}
\label{sec:intro}
The cosmic ray positron flux measured at Earth shows an excess with respect to the secondary positrons, produced by cosmic rays
spallations on the interstellar medium (ISM), at energies above 10 GeV. This was confirmed through a measurement of a rising positron
fraction (defined as the ratio of the $e^+$ to $(e^+ + e^-)$ flux) up to 100 GeV by the {\fontfamily{cmss}\selectfont {PAMELA}} \cite{Adriani:2008zr} and {\fontfamily{cmss}\selectfont {FERMI}}
satellites \cite{FermiLAT:2011ab}. Interestingly, there has been no observed excess in the corresponding cosmic antiproton flux~\cite{Adriani:2010rc, PhysRevLett.117.091103}. Recently, high precision data up to 500 GeV for the positron fraction has been released by the {\fontfamily{cmss}\selectfont{AMS-02}} detector onboard the International
Space Station (ISS) \cite{Accardo:2014lma}. One of the most intriguing features of the {\fontfamily{cmss}\selectfont{AMS-02}} data is that
the positron fraction does not appear to be increasing for energies above $\sim 200$~GeV \cite{Accardo:2014lma}.

\vspace{2mm}
Quantifying the positron excess from the positron fraction typically requires a detailed modelling of the background secondary positrons as
well as the primary electron flux. Regardless of the background modelling uncertainties, it is beyond
reasonable doubt that there is an excess of a primary positron component beyond the secondary astrophysical background from spallations
\cite{Serpico:2008te}, although there is no consensus yet on our understanding of other astrophysical sources of positrons.

\vspace{2mm}
Indeed, purely astrophysical interpretations for a new primary positron component have been proposed such as nearby supernovae
\cite{Fujita:2009wk, Kohri:2015mga} or a population of pulsars localized near the Earth \cite{DiMauro:2014iia, Hooper:2017gtd, Joshi:2017ogv}. In fact, the spectra can also be explained without considering additional sources by realizing the spiral structure of the Milky Way
in 3-D propagation models \cite{Gaggero:2013rya}, and secondary production in shockwaves of supernovae remnants (SNRs) \cite{Blasi:2009hv, Mertsch:2009ph}. Although the pulsar interpretation provides a good fit to the data for reasonable choices of spectral parameters
and spatial distributions (see for example \cite{Cholis:2013psa} or \cite{Boudaud:2014dta} for a recent appraisal), we shall not discuss it
further and refer the interested reader to the literature \cite{Hooper:2008kg, Serpico:2011wg, Linden:2013mqa}. Instead, we argue the case
for an annihilating dark matter (DM) interpretation of the positron excess\footnote{Several interpretations of the PAMELA positron excess
were proposed in the context of annihilating DM in Refs.~\cite{Cholis:2008hb, Cirelli:2008pk, Nomura:2008ru, Harnik:2008uu, Fox:2008kb, Grajek:2008pg} and decaying DM in Refs.~\cite{Ibarra:2008jk, Nardi:2008ix, Arvanitaki:2008hq, Ishiwata:2009vx}. For some recent attempts
at explaining the positron excess with a decaying DM interpretation, see for e.g. \cite{Belotsky:2015gsa, Cheng:2016slx}.}.

\vspace{2mm}
Weakly Interacting Massive Particles (WIMPs) are among the most popular DM candidates, as they can arise `naturally' within models
extending the Standard Model (SM) beyond the electroweak symmetry breaking (EWSB) scale. The typical scenario for production of WIMPs in
the early universe is via the freeze-out mechanism \cite{Kolb:1990vq}. The velocity-averaged $s$-wave annihilation cross-section (only
cross-section henceforth) to achieve a DM thermal relic density consistent with the current value $\Omega_{\text{DM}}h^2 = 0.1193 \pm 0.0014$ \cite{Ade:2015xua}, is computed to be $\left<\sigma v\right>_{\text{TR}} {=} \,\, 2.2 \times 10^{-26}\, \text{cm}^3/\text{s}$ \cite{Steigman:2012nb}, which coincidentally is typical for a ${\sim}$~$\text{TeV}$ scale dark matter particle with SM-like weak couplings.
This coincidence has been dubbed the ``WIMP miracle''.

\vspace{2mm}
Indirect detection signals of WIMP annihilation provide a complementary probe to searches for dark matter at colliders and direct detection
experiments. An excess in the positron flux has long been touted as a `smoking gun' signature for WIMP dark matter annihilations in the Milky
Way galactic halo \cite{Kamionkowski:1990ty}. An analysis of the energy range of the positron excess as well as the observed spectral shape
and normalization allows for a good fit to a ${\sim}$~$\text{TeV}$ scale dark matter particle with a velocity-averaged cross-section $\left<\sigma v\right> \sim 10^{-24} \hyphen 10^{-21} \, \text{cm}^3/\text{s}$ annihilating predominantly into a $W^+ W^-$ pair, or into SM
leptons \cite{Cirelli:2008pk}. Note that this cross-section is ${\sim}$~$10^2 \hyphen 10^5$ times larger than the standard freeze-out
cross-section.

\vspace{2mm}
There are two ways to resolve this discrepancy within the standard WIMP paradigm. The first way is to assume a velocity dependent
Sommerfeld enhancement factor to the cross-section. The approximately inverse dependence of this factor on velocity leaves the early
universe cross-section (during the epoch of freeze-out) essentially unchanged but leads to a large enhancement in the present day
annihilation cross-section due to the low DM velocities ($v\sim10^{-3}c$) that are present in our galaxy today. It has been shown that DM
interactions mediated by SM bosons \cite{Hisano:2004ds, Cirelli:2007xd}, or by a new light mediator \cite{ArkaniHamed:2008qn, Pospelov:2007mp} can give rise to this kind of enhancement. The second approach is to consider astrophysical ``boost factors'' to the annihilation
flux, where overdense regions in the local galactic halo may enhance the DM annihilation rate \cite{Hooper:2008kv} over the canonical
assumption of a smooth dark matter distribution. However in this case, the boost factors have been found to be typically of $\mathcal{O}(10)$ \cite{Pieri:2009je}, which is an insufficient enhancement.

\vspace{2mm}
The annihilating DM interpretation of the positron excess is subject to two types of constraints -- from present day astrophysical
observations and from early universe constraints arising from cosmic microwave background (CMB) observations.

\vspace{2mm}
Astrophysical constraints arise from multi-messenger observations of data from gamma-ray detectors and radio telescopes \cite{Bertone:2008xr, Cirelli:2016mrc, Cirelli:2009dv, Ackermann:2012rg, Ackermann:2013yva, Ackermann:2015zua}. Observations of diffuse gamma rays, gamma rays from dwarf galaxies and radio observations give rise to constraints on annihilating dark matter scenarios. Uncertainty in the intervening astrophysics: DM density profiles \cite{Nesti:2013uwa}, galactic magnetic field modelling \cite{2011AIPC.1381, Jansson:2009ip} and propagation parameters \cite{Genolini:2015cta}, weakens most of these indirect detection constraints allowing some room for a DM interpretation of the positron excess.

\vspace{2mm}
However, early universe constraints from observations of CMB temperature and polarization anisotropy spectra reflecting conditions near the epoch of recombination \cite{Galli:2009zc, Slatyer:2009yq} provide far more robust and stringent constraints. Naive $s$-wave DM cross-sections that explain the positron excess are strongly excluded by these CMB observations. For models with Sommerfeld enhancement, the DM velocity near the epoch of recombination ($v \approx 10^{-8} \, c$) implies that the cross-section is saturated to its maximum value -- greatly exceeding the constraint set by the CMB observations.

\vspace{2mm}
Thus, in light of the CMB constraints, an annihilating DM interpretation of the positron excess has fallen out of favor. In this work we examine the
assumptions behind the CMB constraints and we show that these constraints can be relaxed in the event of a non-thermal production mechanism for
the dark matter particle. We propose a late decaying two-component DM (LD2DM) scenario where the dark sector consists of two semi-degenerate
dark sector particles which we denote as $\chi_1$ (heavier) and $\chi_2$ (lighter)\footnote{Such a class of models has been proposed in the
literature \cite{Cen:2000xv, Abdelqader:2008wa, SanchezSalcedo:2003pb, Peter:2010au, Bell:2010fk} primarily to address the departure, on small scales, of Cold Dark Matter (CDM) simulations from observations \cite{deBlok:2009sp, Klypin:1999uc, Moore:1999nt, Moore:1994yx,  Flores:1994gz, Salucci:2000ps, BoylanKolchin:2011de}.}. Both species are assumed to have simple $s$-wave annihilation cross-sections. The lighter dark matter particle $\chi_2$ is the stable DM candidate that populates the universe
today and annihilates with a large annihilation cross-section to SM particles in order to explain the observed positron excess. The large annihilation
cross-section of $\chi_2$ ensures that it was underproduced in the early universe. The heavy species $\chi_1$ is assumed to have the required
thermal freeze-out cross-section with which it annihilates to a radiation bath. Thus, in the early universe $\chi_1$ plays the role of dark matter with
the correct relic density. The heavy species $\chi_1$ decays on a cosmological timescale through the process,
\begin{equation}
\chi_1 \rightarrow \chi_2 + \phi \nonumber.
\end{equation}
Here, $\phi$ in general stands for one or more light dark sector particles (dark radiation). The late production of $\chi_2$ ensures that its
annihilations to SM particles do not affect the CMB during the epoch of recombination.

\vspace{2mm}
The rest of the paper is organized thus: In Sec.~\ref{sec:DMpositron_b} we examine the constraints from astrophysical and cosmological observations on the best fit dark matter annihilation parameters needed to explain the {\fontfamily{cmss}\selectfont {AMS-02}} positron excess. In Sec.~\ref{sec:problem} we discuss the tension between the standard WIMP
paradigm explanation of the positron excess and (1)~demanding the correct relic density (2)~CMB temperature and polarization anisotropy
constraints on DM annihilation in the early universe. In Sec.~\ref{sec:scenario} we motivate and present the basic LD2DM scenario (along with a toy model that realizes this scenario in Sec.~\ref{sec:toymodel}) as a possible
reconciliation of this conflict, while explaining the observed positron excess. In Sec.~\ref{sec:scenario_b} we discuss how this reconciliation is
achieved and in Sec.~\ref{sec:scenario_c} we examine constraints that arise on the basic LD2DM scenario, from late time energy
depositions (and their effect on the CMB) and from observations of small scale structure. In Sec.~\ref{sec:particle} briefly explore a few phenomenological implications of the toy model. We highlight a few general cosmological features of the LD2DM scenario in Sec.~\ref{sec:features}. Finally, we conclude with an overview of the key results from the paper in Sec.~\ref{sec:discussion}.

In the Appendices we revisit the positron flux data from {\fontfamily{cmss}\selectfont {AMS-02}} and find the best fit dark matter annihilation parameters needed to explain the positron excess. Our results are in general agreement with results found elsewhere in the literature.

\section{Constraints on a dark matter interpretation of the positron excess}
\label{sec:DMpositron_b}

Over the years the cosmic ray (CR) spectra have been measured, to varying degrees of precision, by different experiments. The data for the
positron fraction has been released by {\fontfamily{cmss}\selectfont {PAMELA}} \cite{Adriani:2008zr}, {\fontfamily{cmss}\selectfont {FERMI}} \cite{FermiLAT:2011ab}, and most recently by {\fontfamily{cmss}\selectfont {AMS-02}} \cite{Aguilar:2013qda, Accardo:2014lma}; for the total electron and positron $(e^+ + e^-)$ spectrum by {\fontfamily{cmss}\selectfont {FERMI}} \cite{Ackermann:2010ij}
and {\fontfamily{cmss}\selectfont {AMS-02}} \cite{Aguilar:2014fea}; and for the antiproton spectrum by {\fontfamily{cmss}\selectfont {PAMELA}} \cite{Adriani:2010rc} and {\fontfamily{cmss}\selectfont {AMS-02}} \cite{PhysRevLett.117.091103}. The most intriguing feature
that emerges from the data is that the positron fraction exhibits a steep increase at energies greater than ${\sim}$~$10\textrm{ GeV}$
over the secondary background of positrons expected from CR spallations on the interstellar medium (ISM), whereas no significant excess is
observed for antiprotons. Moreover, measurements of the individual electron ($e^-$) flux by {\fontfamily{cmss}\selectfont {PAMELA}}
\cite{Adriani:2011xv} and {\fontfamily{cmss}\selectfont {FERMI}} prove that the rising positron fraction is indeed due to an excess in the
positron ($e^+$) flux, rather than a decline in the $e^-$ flux. Together, this points towards the presence of a new primary component of
positrons, of an astrophysical origin \cite{Gaggero:2013rya, Fujita:2009wk, Kohri:2015mga, DiMauro:2014iia, Cholis:2013psa, Boudaud:2014dta, Hooper:2008kg, Serpico:2011wg, Linden:2013mqa} or a more exotic origin like DM annihilations.

\begin{figure}[ht!]
                \centering
                \includegraphics[width=0.8\textwidth]{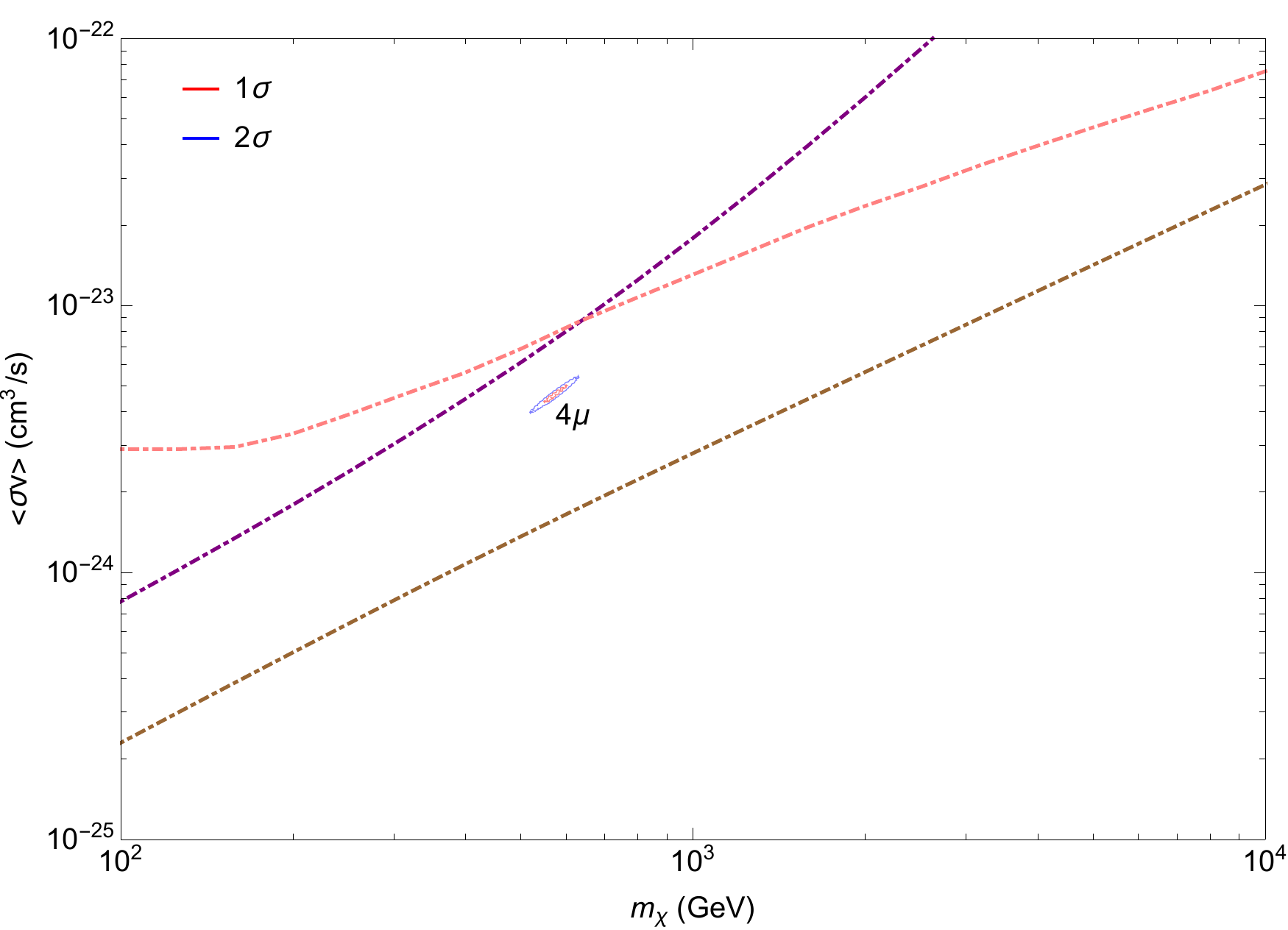}
                \caption{
                The $1\sigma$ and $2\sigma$ contours of the best-fit region for a DM
				explanation of the positron excess are indicated in red and blue respectively for the $4\mu$ annihilation channel.
                Astrophysical constraints on this annihilation channel from {\fontfamily{cmss}\selectfont
               {FERMI}} measurements of gamma rays in dwarf spheroidal
               galaxies (dSphs) (purple, dot-dashed) adopted from
               \cite{Lopez:2015uma, Scaffidi:2016ind}, and diffuse
               gamma constraints (pink, dot-dashed)
               \cite{Papucci:2009gd} are shown. Also plotted are CMB recombination-era
               constraints (brown, dot-dashed), computed using \cite{Slatyer:2015jla, cmb:zz}. NOTE: The CMB constraints depend on the
               assumed cosmological history and do not apply in the LD2DM scenario as discussed in this work.}
                \label{fig:constraints}
\end{figure}

\vspace{2mm}
In this work, we will focus only on the annihilating DM interpretation of the positron excess. In the Appendices we present our calculation of
the best fit dark matter annihilation parameters for various annihilation channels needed to explain the {\fontfamily{cmss}\selectfont {AMS-02}} positron excess. Our results are in general agreement with results found elsewhere in the literature (see for e.g.~\cite{Boudaud:2014dta, Lin:2014vja, Jin:2014ica}). The exact numerical best fit values of DM cross-sections and masses for these annihilation channels 
can be found in Tab.~\ref{table:bestfit} in Appendix~\ref{sec:DMpositron}. Due to the lack of a corresponding excess in the antiproton 
spectrum, we only consider leptophilic annihilation channels, $\mu^ + \mu^- (2\mu)$, $\tau^+ \tau^- (2 \tau)$, $4\mu$ and $4\tau$, in o
our analysis. The $2\mu$ channel is found to have a poor fit to the positron flux whereas the 2$\tau$ and 4$\tau$ channels, although an
excellent fit to the data, are strongly ruled out by gamma ray constraints from either Milky Way dwarf spheroidal galaxies (dSphs) \cite{Lopez:2015uma, Ackermann:2015zua, Abramowski:2014tra} or from the galactic center \cite{Lefranc:2016srp}. Consequently, we 
only focus on the $4\mu$ channel in the rest of this section. Our computation of the best fit $1\sigma$ and $2\sigma$ contours for the 
DM annihilation cross-section and mass (based on our fits to the {\fontfamily{cmss}\selectfont {AMS-02}} positron flux) for the $4\mu$  
channel is shown in Fig.~\ref{fig:constraints}. The annihilating DM interpretation of the positron excess is subject to several astrophysical and
cosmological constraints. We will discuss each of these in turn below.

\vspace{2mm}
\textbf{Astrophysical constraints:}
The annihilation products of the DM should cascade to produce gamma rays as well as undergo acceleration in the galactic magnetic field
(GMF) to produce synchrotron radiation in radio frequencies. Therefore, the annihilating DM interpretation is subject to constraints from multi-messenger analysis from radio and gamma ray spectral data. The radio constraints are subject to large uncertainties inherent in modelling of
the GMF (see \cite{Cirelli:2016mrc} for a recent update), so we only focus on constraints that arise from measurement of the gamma ray
spectrum. There are two types of constraints arising from (1) measurements of the gamma ray flux from Milky Way dSphs (by
{\fontfamily{cmss}\selectfont {Fermi-LAT}}, {\fontfamily{cmss}\selectfont {VERITAS}} \cite{Aliu:2012ga} and {\fontfamily{cmss}\selectfont {HESS}}) and (2) from diffuse gamma ray measurements.

\vspace{2mm}
In Fig.~\ref{fig:constraints}, we also show the  $2\sigma$ upper confidence limit of the DM annihilation cross-section in these channels, using {\fontfamily{cmss}\selectfont {Fermi-LAT}} data for the stacked dSph galaxy gamma ray flux adapted from \cite{Lopez:2015uma, Scaffidi:2016ind}\footnote{These bounds assume that the DM annihilation is mediated by a mediator particle, whose mass is much lighter than the dark matter mass. In the case of the $4\mu$ annihilation channel, as pointed out by \cite{Lopez:2015uma}, radiative corrections must also be included to obtain a competitive constraint \cite{Scaffidi:2016ind}.}. We observe that the $4\mu$ annihilation cross-section needed to explain the positron excess is safe from the dSph bound and is thus a viable annihilation channel for the DM.

\vspace{2mm}
We also include in Fig.~\ref{fig:constraints} the 3$\sigma$ exclusion curve for the $4\mu$ channel from the {\fontfamily{cmss}\selectfont {Fermi}} \textit{diffuse gamma ray measurements} \cite{Papucci:2009gd}, which only has a mild dependence on the Milky Way DM density
profile and is thus more robust than the dSph bound. This bound is conservative, since it attempts to attribute the entire observed diffuse
gamma ray flux to DM annihilations without modelling astrophysical sources. We note from the figure that the cross-section needed to explain
the positron excess in the $4\mu$ channel is safe from this diffuse gamma ray bound\footnote{The bound in \cite{Papucci:2009gd} has
been computed using the preliminary {\fontfamily{cmss}\selectfont {Fermi}} diffuse gamma ray measurements. An updated analysis accounting for the foreground astrophysical diffuse emission has been performed in \cite{Ackermann:2012rg}, however their work does not examine the $4\mu$  channel.}.

\vspace{2mm}
\textbf{Cosmological constraints:}
The most robust and stringent constraints on DM annihilating with large cross-sections comes from the effects of energy deposition into the thermal plasma in the early universe and the consequent effect on the CMB temperature and polarization anisotropies. DM annihilations around the epoch of recombination increase the residual ionization compared to the cosmological ionization history expected in standard $\Lambda$CDM scenarios with non-annihilating DM. For the typical cross-sections under consideration, the increased ionization fraction leaves the peak of the visibility function unchanged, which signifies that the redshift of the last scattering surface (LSS) remains undisturbed. However, the LSS undergoes a broadening \cite{Galli:2009zc, Slatyer:2009yq}, which leaves an imprint on both temperature and polarization spectra of the CMB. The temperature fluctuations on scales smaller than the thickness of the LSS (measured using conformal time) are suppressed, causing a relative attenuation of the power spectrum on these scales, as compared to the standard scenario with non-annihilating DM. This effect alone does not lead to strong bounds on the DM parameters since it is almost completely degenerate with a change in the slope and amplitude of the primordial power spectrum. On the other hand, the effects of broadening of the LSS on polarization spectrum are more distinct. Other than an attenuation effect on small scales, there is also an increase in the amplitude together with a slight shift in the peaks of the E-mode polarization spectrum~\cite{Padmanabhan:2005es}. This allows us to constrain the DM mass and annihilation cross-section. In Fig.~\ref{fig:constraints}, we plot the $2 \sigma$ upper limit on the DM annihilation cross-section in the $4\mu$  channel following the latest CMB anisotropy measurements by {\fontfamily{cmss}\selectfont {Planck}} \cite{Ade:2015xua}, using publicly available recipes \cite{Slatyer:2015jla, cmb:zz}. We see that the CMB constraints are severe and strongly rule out the best-fit DM annihilation masses and cross-sections needed to explain the observed positron excess. We point out here that these CMB bounds are derived under two assumptions (1)~the annihilating DM had a comoving number density ($n$) from around the epoch of recombination till today that varied with redshift $(z)$ as $n \propto (1+z)^3$, i.e. there were no number changing dark matter processes post recombination. This assumption is typical of most DM cosmological histories, including the scenario of WIMPs produced via freeze-out. (2)~ The annihilation cross-section $\langle \sigma v \rangle$ is velocity independent (i.e. $s$-wave scattering cross-section).

\vspace{4mm}
We summarize this section by noting that the {\fontfamily{cmss}\selectfont {AMS-02}} positron excess can \textit{prima facie} be explained by dark matter annihilating to a $4\mu$  final state, with cross-section and mass as given in Tab.~\ref{table:bestfit}. Astrophysical constraints arising from the diffuse gamma ray background and gamma rays from dwarf galaxies cannot rule out these annihilation channels, due to the large astrophysical uncertainties intrinsic to modelling the predicted gamma ray flux. However, under some simple assumptions, CMB constraints prove to be extremely stringent and strongly rule out the best fit cross-sections and masses that we have determined, putting a DM interpretation of the {\fontfamily{cmss}\selectfont {AMS-02}} positron excess in jeopardy.

\section{Beyond the thermal WIMP paradigm}
\label{sec:problem}
As we have seen in the previous section, there are stringent constraints from CMB observations on a dark matter interpretation of the positron
excess. We point out though, that the contours in Fig.~\ref{fig:constraints} have been plotted under the assumptions
that there were no number changing DM interactions in the era leading up to recombination till today and that the value of $\left<\sigma v\right>$
for annihilating DM at recombination is equal to the one that explains the positron excess today. These assumptions implicitly constrain the
cosmological evolution of the DM particle(s).

\vspace{2mm}
For example, within a standard WIMP DM scenario, the DM is a thermal relic that undergoes freeze-out \cite{Kolb:1990vq} with a cross-section
given by $2.2 \times 10^{-26}\, \text{cm}^3/\text{s}$ \cite{Steigman:2012nb}. However, the cross-sections required to explain the
{\fontfamily{cmss}\selectfont {AMS-02}} positron excess are much larger than this value and invoking these cross-sections would give rise to too
low a relic density in the early universe if we assume the standard freeze-out paradigm. One alternative is to posit an astrophysical boost factor from
dark matter clumping on small scales \cite{Hooper:2008kv} in the galactic halo. This would enhance the annihilation rate in the present universe as
compared to the standard expectation from a Navarro-Frenk-White (NFW) \cite{Navarro:1996gj} (or other standard halo) dark matter profile. This explanation is ruled out as the large boost
factors $\mathcal{O}(1000)$ needed to explain the {\fontfamily{cmss}\selectfont {AMS-02}} positron signal are disfavored by numerical
simulations of our galaxy \cite{Pieri:2009je}. Another frequently discussed boost factor is the mechanism of Sommerfeld enhancement~\cite{Hisano:2004ds, Cirelli:2007xd, ArkaniHamed:2008qn, Pospelov:2007mp}, where cross-sections are enhanced at low velocities. The velocity in galactic halos
today is $10^{-3}\,c$, where $c$ is the speed of light, as compared to $0.1\,c$ at the time of freeze-out in the early universe. This implies that the
DM freezes-out with the thermal cross-section, with the correct relic abundance in the early universe, and achieves the larger cross-section
necessary to explain the {\fontfamily{cmss}\selectfont {AMS-02}} positron excess today. However, a Sommerfeld enhancement of the annihilation
cross-section would lead to even more severe bounds from CMB observations since the typical WIMP velocity would be $10^{-8}\,c$ near the
surface of last scattering, thereby saturating the Sommerfeld enhancement factor.

\vspace{2mm}
We summarize the cosmological constraints by noting here a very important point relevant to our current work: \textit{If an $s$-wave or
Sommerfeld enhanced annihilating dark matter species exists before the epoch of recombination, the large annihilation cross-section needed to
explain the} {\fontfamily{cmss}\selectfont {AMS-02}} \textit{positron excess is ruled out by CMB observations.} This bound is
independent of the production mechanism of the DM (see \cite{Baer:2014eja} for a recent review of non-thermal production mechanisms). In the
next section, we propose a scenario that is consistent with the large annihilation cross-sections needed to explain the positron excess today, CMB
observations, and the observed relic abundance.

\section{Late Decaying Two Component Dark Matter (LD2DM) scenario}
\label{sec:scenario}
In order to explain the {\fontfamily{cmss}\selectfont {AMS-02}} positron excess via dark matter annihilations while still being consistent with CMB constraints arising from energy deposition during the epoch of recombination, we adopt an alternate production mechanism scenario for DM which we refer to as late-decaying two-component dark matter (LD2DM). Our framework consists of two dark matter particles, $\chi_1$ and $\chi_2$, characterized by the decay of the heavier particle ($\chi_1$) into the lighter particle ($\chi_2$) on cosmological timescales through the process: $\chi_1 \rightarrow \chi_2 +  \phi $, where $ \phi$ is a light dark sector particle (dark radiation). The decay occurs such that $\chi_2$ is the stable DM that populates the universe today while $\chi_1$ is the one populating the universe during the time of recombination. Hence, by simply attributing the large annihilation cross-section required to explain the positron excess to $\chi_2$ and a significantly smaller one to $\chi_1$, we are able to evade the stringent constraints set by CMB and simultaneously explain the positron excess.

\subsection{Toy model and basic cosmological history}
\label{sec:toymodel}
\vspace{2mm}
For concreteness, we now present a toy model that realizes the basic scenario and explores some of its general features. We emphasize that the scenario is generic and this toy model is
only one of many possible realizations.

\vspace{2mm}
We will assume that $\chi_1$ and $\chi_2$ are both Dirac fermions. We introduce two more fields in our toy model, (1) a dark radiation component ($\phi$), which is a light complex
scalar field that will play the role of the thermal bath for the annihilation of $\chi_1$, and (2) a massive dark force carrier $Z^\prime$ that mediates the annihilation of $\chi_2$ to four
leptons. The interaction terms in the Lagrangian for the effective theory are given by,
\begin{equation}
\mathcal{L}_\textrm{int} = - \frac{1}{\Lambda} \overline{\chi}_1 \chi_1 \phi \phi^* - i g_{\phi} (\phi^* \partial^\mu  \phi - \phi \partial^\mu  \phi^*)  Z^{\prime}_\mu  - i g \overline{\chi}_2 \gamma^\mu \chi_2 Z^{\prime}_\mu   - i g^{\prime} \overline{\psi}_{\ell} \gamma^\mu  \psi_{\ell} Z^{\prime}_{\mu} +  y \overline{\chi}_1  \chi_2 \phi +  \textrm{ h.c.}
\end{equation}
where $\ell$ denotes the heavy SM leptons. We will only consider $\ell$ to be muons since the $4\mu$ annihilation channel is
consistent with gamma ray constraints from dSphs. For $\chi_1$ the dominant annihilation channel is to dark radiation $\phi$, $\phi^*$. The interaction term is written as an effective theory operator with $\Lambda$ denoting the ultraviolet scale of this effective theory. The
couplings of the $Z^{\prime}$ to $\chi_2$, $\ell$ and $\phi$ are denoted by $g$, $g^{\prime}$ and $g_{\phi}$ respectively. The last term
in the interaction Lagrangian violates the $\mathbb{Z}_2$ symmetry of $\chi_1$ and $\chi_2$ and allows $\chi_1$ to decay into $\chi_2$
and $\phi$. We assume that the lifetime of $\chi_1$ is of the order of cosmological timescales, therefore the coupling $y$ must be
extremely small. We take the masses of $\chi_1$ and $\chi_2$ to be denoted by $m_1$ and $m_2$ respectively. The value of $m_2$ is
approximately 570~GeV (taken from Tab.~\ref{table:bestfit} in Appendix~\ref{sec:DMpositron}).

\vspace{2mm}
The dominant annihilation process of $\chi_1$ is $\chi_1 \chi_1 \rightarrow \phi \phi^*$, we denote the ($s$-wave) cross-section for this process to be $\left<\sigma_1 v_1\right>$. We assume the coupling $\Lambda$ is such that $\left<\sigma_1 v_1\right> = \left<\sigma v\right>_{\text{TR}} {=} \,\, 2.2 \times 10^{-26}\, \text{cm}^3/\text{s}$\footnote{We point out that the value of $\Lambda$ needed to achieve the thermal relic cross-section would be close to the mass scale $m_1$ implying a breakdown of the effective theory description of this annihilation process. We take this to imply the existence of a new light mediator for this interaction, but we will avoid introducing the mediator explicitly.}. We assume that the dominant annihilation channel for $\chi_2$ is $\chi_2 \chi_2 \rightarrow Z^{\prime} Z^{\prime} \rightarrow 4\mu$ and we denote the ($s$-wave) cross-section for this process as $\left<\sigma_2 v_2\right>$. We can choose the coupling $g$ such that this cross-section is $\left<\sigma_2 v_2\right> = 4.57\times 10^{-24}\ \textrm{cm}^3/\textrm{s}$\ (again taken from the Tab.~\ref{table:bestfit} in Appendix~\ref{sec:DMpositron}). We will assume a hierarchy $g \gg g^{\prime} \gg g_{\phi}$. This hierarchy of couplings ensures that the annihilation to on-shell $Z^\prime$s which decay to SM leptons is the dominant annihilation channel for $\chi_2$ ($\chi_2 \chi_2 \rightarrow Z^{\prime} Z^{\prime} \rightarrow 4\mu$ as opposed to $\chi_2 \chi_2 \rightarrow Z^{\prime} \rightarrow 2\mu$ or $\chi_2 \chi_2 \rightarrow Z^{\prime} Z^{\prime} \rightarrow 4\phi$).

\vspace{2mm}
The decay lifetime of $\chi_1$ is denoted by $\Gamma^{-1}$. We can estimate the decay width of the heavy dark matter particle $\chi_1$ as,
\begin{equation}\label{eq:lifetime_coupling}
\Gamma \sim \frac{y^2}{4\pi} (m_1-m_2).
\end{equation}

\vspace{2mm}
The large annihilation cross-section of $\chi_2$ would lead to a low abundance for this species in the early universe after its interactions have undergone freeze-out. Since we have assumed that $\chi_1$ has a thermal annihilation cross-section it undergoes freeze-out and would in principle yield the correct dark matter relic density in the early universe, if it were stable. The dark radiation provides a thermal bath for the freeze-out of $\chi_1$ and is kept in equilibrium with the SM thermal bath in the early universe through the interaction with $\mu$ (via the coupling $g_{\phi}$ to the $Z^\prime$).

\vspace{2mm}
In Sec.~\ref{sec:scenario_b} we will examine the evolution of the abundances of both species and demonstrate that the abundance of dark matter (which will be almost entirely in the form of $\chi_2$) in the late universe is correctly reproduced. The annihilation of $\chi_2$ particles in the galactic halo will gives rise to the observed {\fontfamily{cmss}\selectfont {AMS-02}} positron excess. We have assumed that it has precisely the best-fit mass and annihilation cross-section laid out in Tab.~\ref{table:bestfit} (in Appendix~\ref{sec:DMpositron}) to explain the positron excess.

\vspace{2mm}
In Sec.~\ref{sec:scenario_c} we derive the constraints on the mass splitting $(\Delta m \equiv m_1 - m_2)$ and the lifetime of $\chi_1$, $(\Gamma^{-1})$ from cosmology.  We will see that bounds on the lifetime of $\chi_1$ arise from considerations of CMB constraints on energy deposition during the recombination epoch as well as \textit{late time energy deposition} in the cosmic ``dark ages''. Together these constrain $\Gamma^{-1} \gtrsim 10^{13}$~s. We will also see that constraints arising from structure formation limit the typical kick velocity received by daughter particles after decay of the mother particle, and hence lead to a limit on the mass difference between $\chi_1$ and $\chi_2$, $\Delta m \equiv m_1 - m_2 \lesssim 0.1$~GeV.

\subsection{Estimating the cosmological abundances of $\chi_1$ and $\chi_2$}
\label{sec:scenario_b}
\vspace{2mm}
We will first estimate the abundances of the number densities $n_i$ of each dark matter particle ($\chi_1$ and $\chi_2$) at time $t$, by numerically solving the coupled Boltzmann equations:
\beq \label{eq:ndensity}
\begin{aligned}
\frac{d n_1}{d t} + 3 H n_1 &= - \left<\sigma_1 v_1\right> ({n^2_1} - {n_{1 \text{eq}}^2}) - \Gamma n_1,  \\
\frac{d n_2}{d t} + 3 H n_2 &= - \left<\sigma_2 v_2\right> ({n^2_2} - {n_{2 \text{eq}}^2}) + \Gamma n_1,
\end{aligned}
\eeq
where $n_{i \text{eq}}$, $\left<\sigma_i v_i \right>$ $(i=1,2)$ are the equilibrium number densities and annihilation cross-sections for each species and $H$ is the Hubble parameter. Further, following the recipe in \cite{Kolb:1990vq}, we track the number density
in a comoving volume by defining $Y_i (x) = n_i (x)/s(x)$ where $s(x)$ is the entropy density and $x=m_1/T$ is a ``clock''.
We will use the same mass $m \equiv m_1\simeq m_2 = 570$~GeV because as we shall see, structure formation constraints force them to be almost degenerate.

\vspace{2mm}
After these substitutions \Eq{eq:ndensity} takes the form,
\beq \label{eq:ydensity}
\begin{aligned}
\frac{x}{Y_{1 \text{eq}}} \frac{d Y_1}{d x} = - \left(\frac{n_{1 \text{eq}} \left<\sigma_1 v_1\right>}{H(x)}\right)\left[ \left(\frac{Y_1}{Y_{1 \text{eq}}}\right)^2 -1 \right] - \frac{\Gamma}{H(x)} \frac{Y_1}{Y_{1 \text{eq}}}, \\
\frac{x}{Y_{2 \text{eq}}} \frac{d Y_2}{d x} = - \left(\frac{n_{2 \text{eq}} \left<\sigma_2 v_2\right>}{H(x)}\right)\left[ \left(\frac{Y_2}{Y_{2 \text{eq}}}\right)^2 -1 \right] + \frac{\Gamma}{H(x)} \frac{Y_1}{Y_{2 \text{eq}}}.
\end{aligned}
\eeq

\begin{figure*}
\center{
\includegraphics[width=.69\linewidth]{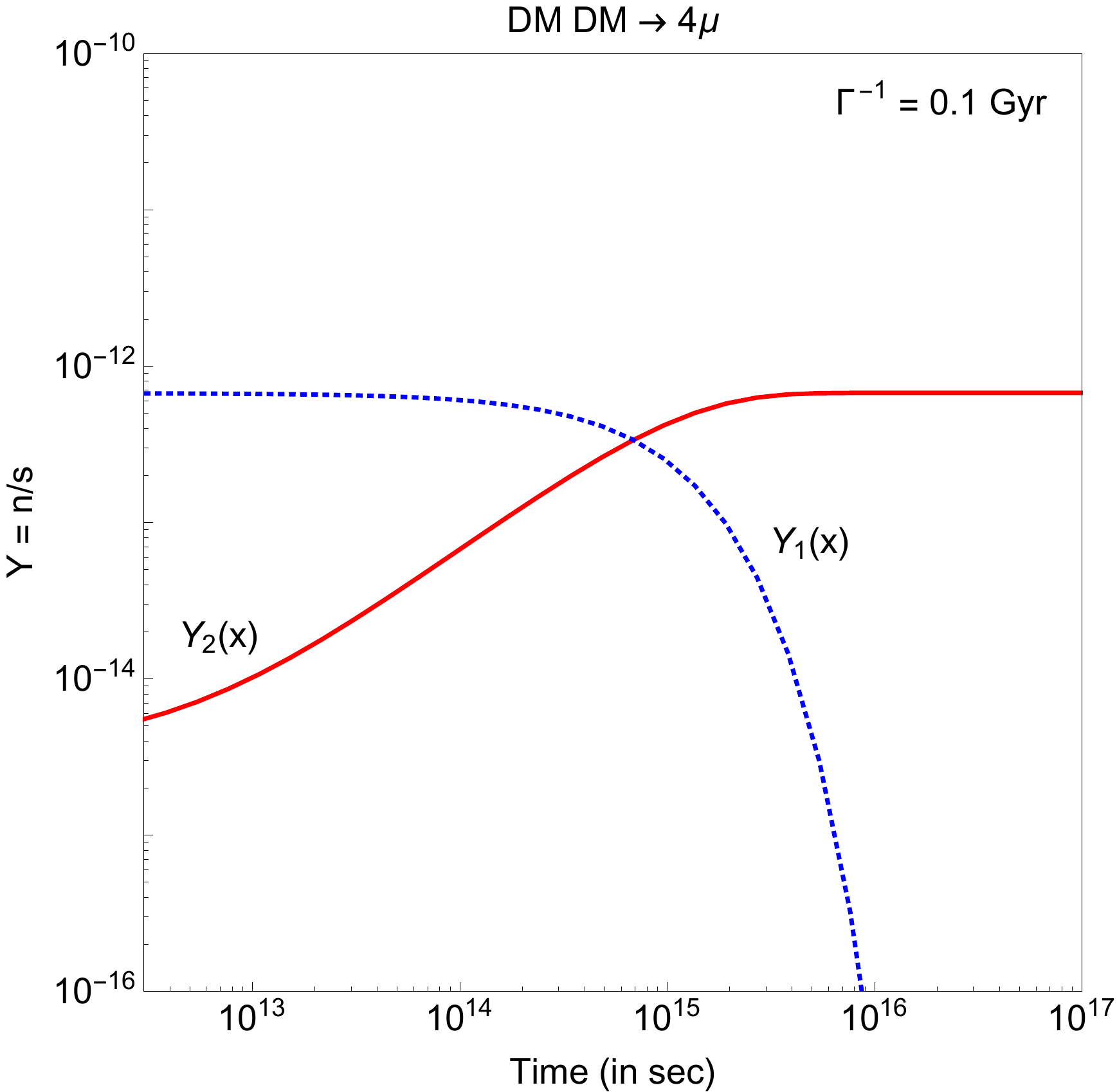}%
}

\caption{The yield $Y\equiv n/s$ of $\chi_1$ and $\chi_2$ as a function of time, assuming a lifetime of $\chi_1$, $\Gamma^{-1}= 0.1$~Gyr. The annihilation cross-sections are
 taken to be $\left<\sigma_1 v_1\right> = 2.2 \times 10^{-26}$~cm$^3$ s$^{-1}$ for annihilation of $\chi_1$ to dark radiation and $\left<\sigma_2 v_2\right> = 4.57 \times 10^{-24}$~cm$^3$ s$^{-1}$ for annihilation of $\chi_2$ to $4\mu$. 
  The late time decay of $\chi_1$ populates $\chi_2$ with the correct dark matter relic abundance. At early times, the abundance of $\chi_2$ is suppressed by a factor of $\mathcal{O}(10^{-2})$ compared to its present day abundance.}
\label{fig:boltzman}
\end{figure*}

\vspace{2mm}
We solve \Eq{eq:ydensity} assuming simple $s$-wave annihilation cross-sections $\left<\sigma_1 v_1\right> = 2.2 \times 10^{-26}$~cm$^3$ s$^{-1}$ and $\left<\sigma_2 v_2\right> = 4.57 \times 10^{-24}$~cm$^3$ s$^{-1}$  for annihilation of $\chi_2$ to $4\mu$. The evolution of the comoving number density in the matter dominated era for a benchmark decay lifetime $\Gamma^{-1} \sim 0.1 \, \text{Gyr}$ is plotted in Fig. \ref{fig:boltzman}. At relatively early times $\chi_1$ and $\chi_2$ are at their would-be freeze-out abundances which are achieved at a temperature $T \sim m/25$, well before recombination. The late decay converts $\chi_1$ into $\chi_2$ species, depleting the abundance of $\chi_1$.

\vspace{2mm}
We can now anticipate how this scenario solves the problems outlined in Sec.~\ref{sec:problem} when trying to reconcile the large DM annihilation cross-section required to explain the positron excess with the naive constraints from CMB observations. The large annihilation cross-section of $\chi_2$ ensures that it has a low abundance in the early universe. This suppressed abundance of $\chi_2$ in the early universe prevents its annihilation products from depositing a significant amount of energy into the thermal plasma during the recombination era. $\chi_2$ eventually attains the correct relic density (and thus forms almost all of the present day dark matter) through the decay of $\chi_1$. Thus, CMB constraints from the broadening of the surface of last scattering should not apply as long as $\Gamma^{-1} \gtrsim 380,000$~years. In the next subsection we will examine the CMB constraints on the late-time annihilation of $\chi_2$ in detail.

\subsection{Constraints on the LD2DM scenario from cosmology}
\label{sec:scenario_c}
There are two types of cosmological constraints that arise on the LD2DM scenario: energy deposition constraints (arising from CMB observations) and structure formation constraints. The first constraint imposes a lower bound on the decay lifetime of $\chi_1$ and the latter constrains the mass splitting between the two DM species. We will discuss each of these constraints in turn.

\subsubsection{Late time energy deposition constraints}
\vspace{2mm}
The late time population of the lighter dark matter particle $\chi_2$ (created by the decays of $\chi_1$) will also undergo annihilations. These annihilations lead to energy deposition into the thermal plasma in the early universe. For lifetimes of $\chi_1$ much smaller than the age of the universe at recombination ($380,000$~years or $10^{13}$~s), the light species $\chi_2$ behaves similarly to a conventional WIMP. Its annihilations in the early universe would lead to energy deposition into the thermal plasma, changing the ionization history of the universe and broadening the surface of last scattering. As discussed in Sec.~\ref{sec:DMpositron_b}, this would predict a distortion of the CMB temperature and polarization anisotropy spectra, which upon comparison with observations, would rule out the large annihilation cross-sections of $\chi_2$ needed to explain the positron excess.

\vspace{2mm}
However for lifetimes of $\chi_1$ greater than the age of the universe at recombination ($\Gamma^{-1} \gtrsim 10^{13}$ s) we expect that the epoch of annihilations of $\chi_2$ will be postponed, leading to late time energy depositions which will alter the ionization history of the universe during the cosmic ``dark ages''. This altered ionization history can also be probed by its influence on the anisotropy spectrum of CMB radiation as it travels from the surface of last scattering towards us. Rather than constraining the annihilation cross-section at a particular epoch (like the recombination era constraints discussed in Sec.~\ref{sec:DMpositron_b}) or by looking at the effect on single observables like the total optical depth (arising from very late decays), Ref.~\cite{Slatyer:2012yq} employs a principal component analysis to provide a bound on the DM annihilation cross-section by considering the complete shape of the ionization history determined by an input cosmological model. Since our scenario has a novel energy deposition history compared to the standard $s$-wave annihilating dark matter, the principal component approach is a useful method to provide a robust bound on the annihilation cross-section of DM from CMB observations.

\vspace{2mm}
We start by discussing the case of a standard thermal relic produced via freeze-out (the conventional WIMP scenario). Post freeze-out, residual dark matter annihilations inject energy into the thermal plasma (or the intergalactic medium (IGM) at late times). These annihilations have an energy injection rate per comoving volume given by,
\begin{align}
\left . \frac{dE}{dt dV}\left( z \right)  \right |_\mathrm{injected}  = \kappa \left ( 2 m_{\mathrm{DM}} c^2 \right )  {n}^2(z) \langle \sigma v \rangle,
\label{eq:einjected}
 \end{align}
 where $\kappa =1/2$ for majorana-like DM and $\kappa = 1/4$ for Dirac-like DM. Here, $z$ is the redshift at which the annihilation occurs, $m_{\mathrm{DM}}$ is the mass of the dark matter particle, $n(z)$ is the number density of DM particles and $\langle \sigma v \rangle$ is the velocity-averaged annihilation cross-section.

\vspace{2mm}
For a standard thermal relic, its cosmological abundance at a redshift $z$ is given by,
\begin{align}
 n_\textrm{TR}(z) =\frac{ \rho_c}{m_\textrm{DM}} \Omega_{\textrm{DM}} (1+z)^3, \label{eq:nthermrelic}
\end{align}
where $\rho_c$ is the critical density and $\Omega_{\textrm{DM}} \sim 0.2$ is the present day dark matter energy density fraction.

\vspace{2mm}
This \textit{energy injection} via dark matter annihilations does not however give the correct \textit{energy deposition} history. To get the correct energy deposition history one needs to include a redshift dependent efficiency factor $f(z)$, defined such that,
\begin{align}
\left(\frac{dE}{dt dV} \right)_\mathrm{deposited} & = f(z)  \left(\frac{dE}{dt dV} \right)_\mathrm{injected}.
\label{eq:fzdef}
\end{align}

\vspace{2mm}
The detailed derivation of the factor $f(z)$ has been shown in  \cite{Slatyer:2012yq}. We highlight here some key factors that determine $f(z)$ and write down the final expression without repeating the derivation.

\begin{enumerate}
\item In Ref.~\cite{Slatyer:2012yq} it was shown that energy $E$ injected in photons and positrons at an early redshift $z_{\textrm{inj}}$ will only be deposited as ionizing radiation at a later redshift $z_{\textrm{dep}}$. The author provided tables of the fractional energy deposited as a function of initial energy for various injection and deposition redshifts, $T(z_{\textrm{dep}},z_{\textrm{inj}},E)d\textrm{ln}(z_{\textrm{dep}})$\footnote{These energy deposition fractions were updated in Ref.~\cite{Slatyer:2015kla} to correctly account for different energy deposition channels (ionization, heating and excitation of the gas and low energy photons which would distort the CMB spectrum). Our interest here is only the ionization channels of Ref.~\cite{Slatyer:2015kla}, which we use to compute the corrected ionizing energy deposition fractions.}, for various electromagnetic species (photons and electron-positron pairs).
\item The energy injection formula we are considering only includes annihilations from the smooth, isotropic component of dark matter. One needs to include a boost factor $\mathcal{B}(z)$ due to halo formation at late times, which enhances the dark matter annihilation probability as compared to the smooth component alone.
\item An effect that is new in the LD2DM scenario as compared to previous studies of energy deposition constraints, is that the number density of annihilating dark matter particles in our scenario differs from the standard thermal relic scenario\footnote{In Ref.~\cite{Slatyer:2012yq}, a somewhat similar situation was considered in the case of oscillating asymmetric dark matter.}. Recall that the ``active species'', $\chi_2$ is produced initially with a very low abundance due to freeze-out and is later repopulated via the decay of $\chi_1$.
\end{enumerate}

\vspace{2mm}
Thus in the LD2DM scenario, a clearer definition of $f(z)$ is provided by,
\begin{align}
\left(\frac{dE}{dt dV} \right)_\mathrm{deposited, smooth + halos, LD2DM} & = f(z)  \left(\frac{dE}{dt dV} \right)_\mathrm{injected, smooth, TR}.
\label{eq:fzdefld2dm}
\end{align}
Note that as a normalization convention, we choose the injected energy to originate from a smooth thermal relic, however when calculating the deposited ionizing energy (which affects CMB observables) we are considering all of the factors listed above.

\vspace{2mm}
The resulting formula for $f(z)$, after taking these effects into account, is given below,
\begin{align}
f(z)    = \frac{H(z)}{(1+z)^3 } \times \frac{\displaystyle \sum_\mathrm{species (s)} \int \frac{(1+z')^2 dz'}{H(z')}\left( 1+ \mathcal{B}(z') \right) \left(\frac{n_2^2(z')}{n_\textrm{TR}^2(z')}\right) \int T_s(z',z,E_s) E_s \frac{d\bar{N_s}}{dE_s}dE_s}{\displaystyle \sum_\mathrm{species (s)} \int E_s \frac{d\bar{N_s}}{dE_s} dE_s}. \label{eq:annf}
\end{align}
Here, $H(z)$ is the Hubble parameter and $\frac{d\bar{N_s}}{dE_s}$ is the spectrum of particles of type $s$ ($e^+e^-$ pairs or photons) per annihilation. The boost factor due to halos is given by \cite{Poulin:2015pna},
\begin{align}
 \mathcal{B}(z) =\frac{\bar{f}_h}{(1+z)^3}  \mathrm{erfc}\left(\frac{1+z}{1+z_h}\right), \label{eq:haloboostf}
\end{align}
where $z_h$ is the redshift at which halos start to contribute an enhancement to the DM annihilations and $f_h$ is proportional to a halo ``concentration function''. Following Ref.~\cite{Poulin:2015pna}, we take $\bar{f}_h = 10^9$ and $z_h =20$. We take the comoving number density of $\chi_2$ ($n_2$) as computed in Sec. \ref{sec:scenario_b} and $n_{\textrm{TR}}$ is the comoving number density of a thermal relic. Note that the factor $\left(\frac{n_2^2(z')}{n_\textrm{TR}^2(z')}\right)$ is unique to the LD2DM scenario, and it arises because of the alternate cosmological history of $\chi_2$. We were able to numerically compute the energy deposition efficiency factor $f(z)$ using the Mathematica~\cite{Wolfram2010} tools provided by \cite{Slatyer:2012yq}. For each final state, the energy injection spectrum $\frac{d\bar{N_s}}{dE_s}$ into photons, electrons and positrons is different. We used the {\fontfamily{cmss}\selectfont{PPPC4DMID}}\footnote{\href{http://www.marcocirelli.net/PPPC4DMID.html}{http://www.marcocirelli.net/PPPC4DMID.html}} code~\cite{Cirelli:2010xx, Ciafaloni:2010ti} to obtain the spectrum for the $e^+e^-$ pairs and photons for the $4\mu$ annihilation channel. We then computed $f(z)$ for each annihilation channel for different values of the heavy species decay lifetime $\Gamma^{-1}$ (which controls the abundance $n_2(z)$ of $\chi_2$).

\vspace{2mm}
Once we obtained the efficiency factor $f(z)$, we were able to compute the ionizing energy deposition history for various annihilation cross-sections $\langle \sigma_2 v_2 \rangle$. Next, we examined CMB constraints on this energy deposition history. Ref.~\cite{Slatyer:2012yq} used a principal component analysis (PCA), first developed in~\cite{Finkbeiner:2012aa}, to constrain the energy deposition history and consequently the annihilation cross-section. We followed the same approach to constrain the annihilation cross-section $\langle \sigma_2 v_2 \rangle$. Using a modified version of the Mathematica code provided by \cite{Slatyer:2012yq}, we computed the upper bound on the annihilation cross-section $\langle \sigma_2 v_2 \rangle$ for different values of the lifetime of $\chi_1$.

\vspace{2mm}
We note here a subtlety of our analysis: for compatibility with the PCA approach we assumed that the cross-section $\langle \sigma_2 v_2 \rangle$ only enters the energy deposition history \textit{linearly}. However, this is not really the case in the LD2DM scenario since the abundance $n_2(z)$ at early times is also controlled by $\langle \sigma_2 v_2 \rangle$. We will ignore this subtlety by using a fixed cross-section in the calculation of $n_2(z)$ (We choose the best fit cross-sections needed to explain the {\fontfamily{cmss}\selectfont {AMS-02}} positron excess). Another issue is that the PCA approach that we consider does not provide the most reliable bound on energy deposition when the lifetime of $\chi_1$ is of the order of the age of the universe today ($\Gamma^{-1}\geq10^{16}$~s). This is because for such late time energy depositions astrophysical processes contribute to ionization of the intergalactic medium (IGM) and this contribution must be accounted for when setting a constraint on the DM annihilation cross-section. For such late time decays, we expect the bounds on the annihilation cross-sections from other sources, such as the heating of the IGM to also be important \cite{Diamanti:2013bia, Liu:2016cnk}.

\begin{figure*}
 \center{
 \label{sfig:testa}  \includegraphics[width=.69\linewidth]{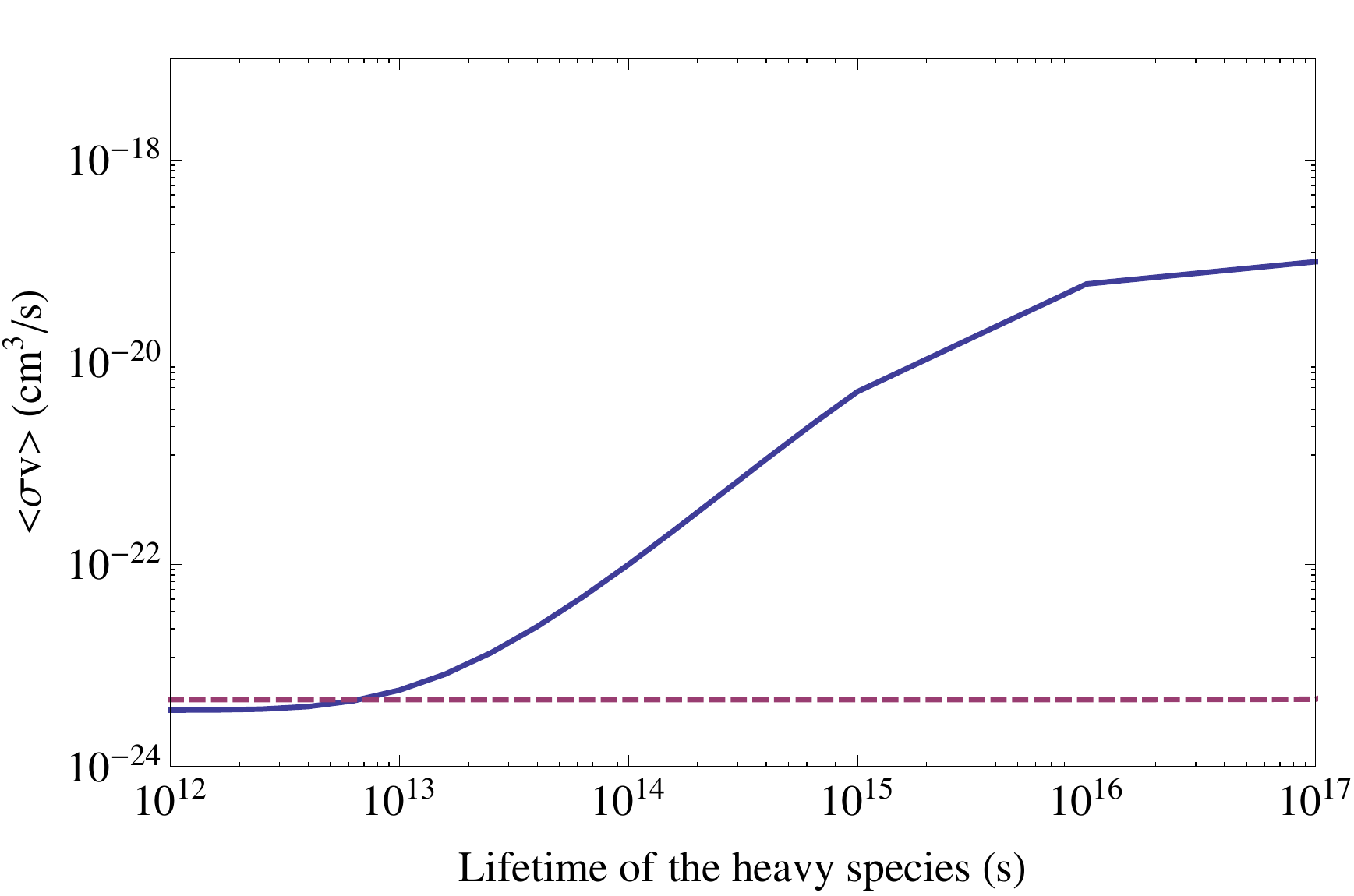}%
 }
\caption{Upper bound on the annihilation cross-section (solid blue curve) of $\chi_2$ to $4\mu$ 
from the energy deposition history constraints arising from observations of the CMB. Also shown is the required annihilation cross-section to explain the {\fontfamily{cmss}\selectfont {AMS-02}} positron excess (dashed magenta line).}
\label{sfig:energydepbound}
\end{figure*}

\vspace{2mm}
Our resulting upper bounds on $\langle \sigma_2 v_2 \rangle$ as a function of the lifetime $\Gamma^{-1}$ are shown in Fig.~\ref{sfig:energydepbound}. We also plot the required annihilation cross-section necessary to explain the {\fontfamily{cmss}\selectfont {AMS-02}} positron excess for comparison. For short lifetimes of $\chi_1$ (smaller than the time of recombination, $\Gamma^{-1} < 10^{13}$~s), we see that the bound saturates to a constant value as expected in the case of a standard WIMP type scenario. In this case, the bound primarily arises due to the effect on CMB anisotropies from the broadening of the surface of last scattering. As we have already seen in Sec.~\ref{sec:DMpositron_b}, the resulting constraint rules out the annihilation cross-sections needed to explain the positron excess in the $4\mu$ channel. However, for lifetimes greater than $\sim 10^{13}$~s, we find that the late time energy deposition constraints are less stringent and the required annihilation cross-sections to explain the {\fontfamily{cmss}\selectfont {AMS-02}} positron excess are consistent with the resulting CMB bounds.

\subsubsection{Structure formation constraints}
\vspace{2mm}
In the LD2DM scenario the daughter particles $(\chi_2)$ receive a kick velocity $v_k$ in the frame of the mother particles ($\chi_1$) which is proportional to the mass difference $\Delta m \equiv m_1-m_2$ between $\chi_1$ and $\chi_2$,
\begin{equation}
v_k=\frac{\Delta m}{m_1}c,
\end{equation}
where $c$ is the speed of light. The resultant kick velocity strongly impacts the structure formation process in the universe. The free 
streaming length of the dark matter suddenly increases after the decay of $\chi_1$, suppressing all structure below the enhanced free 
streaming length scale. Using Lyman-$\alpha$ forest data, Ref.~\cite{Wang:2013rha} finds that the kick velocity, $v_k$, received by the 
daughter particles is restricted to be less than 40~km/s for $\Gamma^{-1} < 10$ Gyr. Thus the strong constraints on $v_k$ essentially 
restricts the mass of $\chi_1$ to be nearly degenerate with $\chi_2$, with $\Delta m \lesssim 0.1$~GeV.

\section{Particle physics features of the toy model}
\label{sec:particle}
\vspace{2mm}
In this section we discuss some potential phenomenological implications of the LD2DM scenario. We note here that some of these features are generic to any model that realizes the
LD2DM scenario while some of the signatures are specific to the toy model presented in Sec.~\ref{sec:toymodel}. The toy model makes some minimal assumptions for economy of
particle content:
\begin{itemize}
\item One key assumption is that the particle that makes up the dark sector thermal bath that $\chi_1$ annihilates to while undergoing freeze-out is the same as the particle $\phi$
involved in the decay $\chi_1 \rightarrow \chi_2 + \phi$.
\item The second minimal assumption is that the mediator $Z^\prime$ that is responsible for the annihilation of $\chi_2$ to 4 leptons is also responsible for keeping the dark radiation
bath in kinetic equilibrium with the visible sector (at least in the early universe).
\end{itemize}
These assumptions can be relaxed without affecting our basic scenario. In fact, it is not necessary that $\chi_1$ is produced via freeze-out as long as it has the correct relic density
before it starts decaying to $\chi_2$.

\vspace{2mm}
The particles introduced in the toy model could potentially gives rise to (or be constrained by) cosmological observations and signatures at terrestrial experiments. We will briefly address these below:
\begin{itemize}
\item Since the $Z^\prime$ couples to muons, the coupling $g^\prime$ would be strongly constrained by muon $g-2$ measurements and must therefore be small.
\item The dark radiation contributes to the number of effective neutrinos $\Delta N_{\textrm{eff}}$ measured by CMB experiments, however the exact contribution depends on the temperature of the radiation bath relative to the visible sector bath at the epoch of recombination. This temperature is controlled by the efficiency of the $\phi-\mu$ scattering process which also depends on the $Z^\prime$ coupling to muons. If this coupling is small, the dark radiation bath decouples early and has a low temperature as compared to CMB photons which are heated by annihilating particles such as $e^+e^-$ pairs. This would result in a negligible contribution to $\Delta N_{\textrm{eff}}$.
\item We could also consider $\phi$ to be a visible sector particle in which case the only candidates are photons and neutrinos, due to the small mass splitting between $\chi_1$ and $\chi_2$ (we would have to change the spins of $\chi_1$ and $\chi_2$ in our toy model in the latter case). Photons would again be subject to energy deposition constraints from decaying dark matter and are hence disfavored~\cite{Slatyer:2012yq,Oldengott:2016yjc,Slatyer:2016qyl}. However, the case where $\phi$ is a neutrino is interesting and could lead to features in the Cosmic Neutrino Background at energies of \linebreak $\mathcal{O} (\Delta m /(1+z_d))$, where $z_d$ is the characteristic redshift for the decay of $\chi_1$.
\end{itemize}

\vspace{2mm}
Apart from these experimental and observational issues, we run into two main fine-tuning problems when trying to build a detailed particle physics model that realizes the basic LD2DM scenario, as can be seen in the parameters of our toy model.
\begin{itemize}
\item If we choose the lifetime to be of the order of $0.1$~Gyr ($\sim$ $10^{16}$~s), for a mass-splitting of $\mathcal{O}(0.1)$~GeV, we find that we need to choose $y \sim 10^{-11}$ using \Eq{eq:lifetime_coupling} . Hence, this coupling is highly tuned.
\item The mass splitting between $\chi_1$ and $\chi_2$ is extremely small, with $\epsilon \equiv \frac{\Delta m}{m_1} \sim 10^{-4}$.
\end{itemize}
The first fine tuning problem could potentially be solved by positing a $\mathbb{Z}_2$ symmetry for $\chi_1$ that is broken spontaneously but communicated through loops or via highly suppressed mass scales. The second fine tuning is more problematic because $\chi_1$ and $\chi_2$ have different effective couplings to SM muons and hence they receive different quantum corrections to their masses.

\vspace{2mm}
We emphasize here that in this work we have merely highlighted some interesting implications of models that incorporate the basic LD2DM scenario. We leave a detailed development of the toy model and various extensions, along with a full analysis of their parameter space and particle phenomenology to future work.

\section{Possible cosmological features of the toy model}
\label{sec:features}
In this section we will discuss several small scale structure and cosmological problems and show that the toy model discussed in Sec.~\ref{sec:toymodel} has the potential to resolve these problems.

\subsection{Impact on small scale structure}
\vspace{2mm}
The cold dark matter (CDM) paradigm has been very successful in explaining the large scale structure of the universe but on small scales there are hints from observations of dwarf galaxies of the Milky Way (MW) and M31 that something may be lacking in this description. Precision observations of dwarf galaxies show DM distributions with cores, in contrast to cusps predicted by CDM simulations (``cusp-core'' problem) \cite{Oh:2010ea}. It has also been shown that the most massive subhalos in CDM simulations of MW size halos have higher central densities than those observed in the brightest satellites of the MW (``too big to fail problem'')~\cite{BoylanKolchin:2011de}. Simulations also predict numerous dwarf satellites that are not seen (``Missing Satellite Problem'')~\cite{Klypin:1999uc, Moore:1999nt}. The ``missing satellite problem'' could potentially be explained by our failure to detect ultra faint dwarf satellites~\cite{Tollerud:2008ze} in the MW. However, several potential solutions to these small scale structure problems that rely on the inclusion of baryonic physics in DM halo dynamics have also been proposed. For example, the ``missing satellite problem'' could alternatively be explained by baryonic processes which inhibit the accretion of gas in low mass subhalos~\cite{Bullock:2000wn, Benson:2001at}. Similarly,~\cite{Governato:2012fa} found that baryonic processes like episodic supernova feedback can potentially flatten the cuspy DM profile at the center of a halo, but it is unlikely that the ``too big to fail'' problem can be resolved by such processes alone \cite{GarrisonKimmel:2013aq}. It is also worthwhile to consider the predictions of alternative dark matter physics for these issues (see \cite{Weinberg:2013aya} for review).

\vspace{2mm}
Models which subscribe to the LD2DM paradigm consist of two approximately degenerate DM particles $\chi_1$ and $\chi_2$. The heavy species $\chi_1$ would undergo decay on a time scale (possibly) of the order of the age of the universe producing $\chi_2$ with a non-relativistic kick velocity $v_k$. As a result, dark matter halos which have a maximum circular velocity, $V_{max}$, close to the value of the kick velocity are disrupted once significant decay has occurred. From the Lyman-$\alpha$ constraints discussed in Sec.~\ref{sec:scenario_c}, we find that the allowed values of $v_k$ for $\Gamma^{-1}$ less than the age of the universe are of the order of $10$~km/s. This is around the order of the typical $V_{max}$ values found in dwarf galaxies. Hence, the effect of a kick velocity imparted during decay of the heavy DM species would be most readily visible through its impact on our galactic halo substructures. Subhalos which have $V_{max}$ less than $v_k$ will almost completely disintegrate after all the mother particles have decayed. Simulations of a MW sized halo using models where an unstable dark matter particle decays into a stable non-interacting dark matter particle were performed by \cite{Wang:2014ina}, and they found a significant reduction of subhalo count for $\Gamma^{-1}=1$~Gyr and $10$~Gyr and $v_k=20$~km/s. This indicates that decaying dark matter has the potential to solve the ``Missing Satellite Problem'' for an appropriate choice of $v_k$ (or a corresponding choice of the mass splitting). For $V_{max} \gg v_k$ almost no impact should be felt by the subhalos. However, subhalos which have $V_{max}$ slightly greater than $v_k$ will experience expansion due to the sudden injection of energy into the system through decays. The subhalos will also suffer from some fraction of mass loss as some initially loosely bound particles will decay to particles that are no longer bound to the subhalo. The net effect is that there will be a reduced central halo density~\cite{Peter:2010au}. However, this effect is only dominant for values of $V_{max}$ close to $v_k$. Ref.~\cite{Wang:2014ina} found that $\Gamma^{-1}=10$~Gyr and $v_k=20$~km/s sufficiently drove down the central density of the largest subhalos in the simulations, which they concluded would resolve the ``too big to fail'' problem, although the density profiles still remained cuspy.

\subsection{Impact on Hubble tension and $\sigma_8$ tension}
\vspace{2mm}
Recently, it has been observed that the Planck best fit parameters assuming the concordant $\Lambda$CDM cosmological model~\citep{Ade:2015xua} prefer $\sigma_8$ and $H_0$ values which are different from the values that are inferred from direct observations. In particular, Planck prefers a value of $H_0$ which is smaller than the value from low-redshift measurements of supernovae~\cite{Verde:2013wza, Zhang:2017aqn}, while the preferred $\sigma_8$ value is larger than the directly measured value extracted from observations of large scale structure~\cite{Battye:2014qga}. While it is premature to rule out the contribution of underestimated systematic errors on the direct measurements in these discrepancies, they still motivate theories beyond the standard $\Lambda$CDM scenario. The toy model we considered in Sec.~\ref{sec:particle} allows room for additional interactions between a dark radiation component ($\phi$) and a primordial dark matter particle ($\chi_1$), which under certain conditions could potentially resolve the tensions in the $H_0$ and $\sigma_8$ measurements simultaneously. We will elaborate on this resolution below.

\vspace{2mm}
In the early universe, when all the neutrinos are relativistic, the effective number of neutrinos, $N_{\textrm{eff}}$, can be defined as a proxy variable to describe the total radiation energy density ($\rho_r$) in terms of the photon energy density $\rho_{\gamma}$:
\begin{equation}
\rho_r=\rho_{\gamma}+\rho_{\nu} + \rho_{\textrm{DR}} \equiv\bigg[1+\frac{7}{8}\bigg(\frac{4}{11}\bigg)^{4/3}N_{\textrm{eff}}\bigg]\rho_{\gamma},
\end{equation}
where, $\rho_{\nu}$ and $\rho_{\textrm{DR}}$ are the densities of neutrinos and a putative dark radiation component respectively. In the standard $\Lambda$CDM scenario where the dark radiation component is absent, $N_{\textrm{eff}}$ has a nominal value of 3.045~\cite{Mangano:2005cc,deSalas:2016ztq}. The addition of dark radiation in our toy model would increase $N_{\textrm{eff}}$ as,
\begin{equation}
\Delta N_{\textrm{eff}}= N_{\textrm{DR}}\bigg(\frac{T_{\textrm{DR}}}{T_{\nu}}\bigg)^4\times\Bigg\{^{8/7 \ \textrm{(bosonic DR)}}_{1 \ \textrm{(fermionic DR)}}.
\end{equation}
Here, $T_{\textrm{DR}}$ and $T_{\nu}$ denote the temperatures (at recombination) of dark radiation and neutrinos respectively. Thus, an increase in dark radiation would lead to an increase in the parameter $N_{\textrm{eff}}$. Such an increase in $N_{\textrm{eff}}$ due to dark radiation can be compatible with the observed CMB data provided that the characteristic redshifts of matter/radiation equality and matter/cosmological constant equality remain unchanged. This can be accommodated by an appropriately larger value of $H_0$. A recent analysis found that an extra dark radiation component with $\Delta N_{\textrm{eff}}\approx 0.4-1$ can in fact reconcile the measurements from HST with Planck data~\cite{Riess:2016jrr}. This is also compatible with big bang nucleosynthesis constraints on extra relativistic degree of freedom, $\Delta N_{\textrm{eff}}\leq 1$ at 95\% C.L (see~\cite{2011PhLB..701..296M}). Thus, it is clear that the data is consistent with the inclusion of an extra dark radiation component (as in our toy model) in the universe and this would also allow for a resolution to the $H_0$ tension.

\vspace{2mm}
Although an increase in $N_{\textrm{eff}}$ alleviates the tension in $H_0$, it would also exacerbate the $\sigma_8$ tension by increasing the best fit value obtained from CMB data~\cite{Ade:2015xua}. Such a conflict could be resolved by introducing a drag force between dark radiation and dark matter for which the interaction rate scales with the Hubble parameter during the radiation dominated era~\cite{Lesgourgues:2015wza}. This coupling essentially restricts the growth of dark matter over-densities at all scales within the Hubble horizon during the radiation dominated era. This also ensures that the scales crossing the Hubble horizon during the matter dominated era will remain unaffected as the Hubble rate is larger than the interaction rate in this epoch. Hence, the power spectrum on relatively small scales (including $8/h$~Mpc) could potentially be suppressed, which would reconcile the inferred value of $\sigma_8$ extracted from the normalization of the CMB power spectrum with the directly obtained value from observations of large scale structure. Thus, introducing a drag force of this type between the primordial dark matter $\chi_1$ and dark radiation in our toy model could potentially resolve the $\sigma_8$ tension as well.

\section{Summary and Discussion}
\label{sec:discussion}
\vspace{2mm}
The {\fontfamily{cmss}\selectfont {AMS-02}} positron excess continues to be a puzzle several years since the original discovery. In this
work, we have explored an annihilating DM interpretation of the positron excess. We compared the best-fit parameters needed to explain the
excess with constraints arising from the observed gamma ray flux from dwarf galaxies and diffuse gamma ray measurements. We found
that among the four leptophilic DM annihilation channels we considered, i.e. 2$\mu$, 2$\tau$, 4$\mu$ and 4$\tau$, only the $4\mu$
channel was still allowed by these constraints. The most stringent constraint on such a dark matter candidate arises from observations of
the CMB. As we have argued, if an $s$-wave annihilating dark matter species exists before the epoch of recombination, it would deposit
energy into the thermal plasma in the early universe with observable effects on the CMB temperature and polarization power spectra. The
large cross-section needed to explain the positron excess would certainly be ruled out, putting an annihilating DM interpretation of the
positron excess in serious jeopardy.

\vspace{2mm}
The main feature of this work was our proposal of a novel paradigm, which we dubbed the late-decaying 2-component dark matter (LD2DM)
scenario. In this scenario, there is a heavy dark matter species, $\chi_1$, that annihilates with a low (thermal relic) cross-section, while
decaying on cosmological time scales to a more strongly annihilating lighter DM species, $\chi_2$. We posit that $\chi_2$ plays the role of
DM in the present universe and its annihilations give rise to the observed positron excess. For concreteness, we presented a toy particle physics model that realizes the basic LD2DM scenario.

\vspace{2mm}
We showed, through an explicit calculation of the yields, that the LD2DM scenario can give rise to the correct relic abundance of DM in the present universe while evading CMB constraints on
energy deposition during the epoch of recombination. We then proceeded to examine constraints that arise on the LD2DM scenario from cosmology, specifically (1) constraints arising from
\textit{late-time energy depositions} during the cosmic dark ages and their effect on the CMB and (2) structure formation constraints on
decaying dark matter. We found that these constraints force us to consider $\chi_1$ to be a long lived DM candidate with a lifetime greater than $\sim 10^{13}$~s and a very small mass splitting between $\chi_1$ and $\chi_2$.

\vspace{2mm}
We then discussed possible phenomenological implications of our toy model: potential observations, constraints and fine-tuning issues in this model. We also explored several possible cosmological implications of this toy model,
such as the possibility to solve certain small scale structure problems and the potential to resolve the tension between the values of $\sigma_8$ and $H_0$ extracted from CMB observations with those from low-redshift measurements.

\vspace{3mm}
In this work we have left open several avenues that could be pursued if the basic LD2DM paradigm is taken seriously as an explanation of the
positron excess. We highlight below some directions for future work:

\vspace{2mm}
\textbf{DM and the positron excess:}
The very first question that needs to addressed is whether the observed positron excess will continue to favor a dark matter interpretation once more data is collected. The {\fontfamily{cmss}\selectfont {AMS-02}} detector is designed to continue collecting data through 2020, so future releases of data will improve precision in the high energy bins and
reveal the shape of the positron fraction (flux) that may help in distinguishing between astrophysical and DM interpretations of the excess \cite{AMS:2015icrc}. Also, as advocated by \cite{Cirelli:2015gux}, determining the energy-dependent anisotropy of the incoming positrons may be an indicator of the nature of its source: a high anisotropy would favor a single-
source hypothesis while a DM interpretation is better served by an isotropic positron source.

\textbf{Particle physics models of the LD2DM scenario:} We leave a detailed development of the toy model and various extensions, along with a full analysis of the parameter space and
particle phenomenology to future work. One potentially promising phenomenological avenue is to look at scenarios where $\chi_1$ decays to $\chi_2$ and a neutrino. This could lead
to observable signatures in the Cosmic Neutrino Background. On the model building side, it would be interesting to see whether the fine tunings in our toy model can be explained by
embedding it in a UV completion.

\textbf{21cm astronomy and Epoch of Reionization:}
It is interesting to consider the LD2DM scenario for a decay lifetime of $\chi_1$ of the order of the age of the universe. $\chi_2$ has a significantly larger cross-section compared to
the vanilla WIMP particle and hence can have a substantial contribution in saturating the optical depth to reionization. On top of that, $\chi_2$ also has a varied abundance history which
would cause its energy deposition record to be different from that of conventional $\Lambda$CDM cosmologies. Hence, we expect the LD2DM scenario to leave a distinct imprint on the
reionization history of the universe. The observations of redshifted 21cm radiation from the Epoch of Reionization (EoR) will be essential in providing constraints on the contribution to
reionization process from various sources including DM annihilation. The ongoing experiments like HERA \cite{Pober:2013jna} are expected to provide us with exciting information like
global history of EoR progression, but they will only be limited up to $z \approx 11$. In the future, experiments like SKA \cite{Mesinger:2015sha} will have much better sensitivity,
redshift coverage (upto $z \approx 30$) and resolution which will help shed light on the dark ages. This will also play a major role in probing alternate reionization histories as maybe the
case in the LD2DM scenario.

\newpage
\section*{Acknowledgments}
We are grateful to Andre Scaffidi for providing the constraint plot for $4\mu$ channel after including radiative corrections using the newest
{\fontfamily{cmss}\selectfont{Fermi-LAT}} likelihoods from \texttt{Pass 8} data. We thank Marco Cirelli, Basudeb Dasgupta, Fabio Iocco,
and Rishi Khatri for useful discussions. We especially thank Tracy Slatyer for helpful correspondence on the PCA constraints for late time dark
matter annihilations. JB would like to acknowledge the hospitality of Mainz Institute for Theoretical Physics (MITP) where a part of this work
was completed. The work of VR was supported by an IRCC, IIT-B seed grant.

\vspace{2mm}
Finally, we thank the three anonymous referees for their critique and helpful suggestions.

\appendix

\section{Annihilating dark matter and the positron excess}
\label{sec:DMpositron}
\vspace{2mm}
For the first time, the {\fontfamily{cmss}\selectfont{AMS-02}} collaboration has released data for the electron and positron fluxes separately \cite{Aguilar:2014mma}, along with the updated $(e^+ + e^-)$ flux~\cite{Aguilar:2014fea} and positron fraction data. This high-precision data for the cosmic positron flux (see black data points in Fig.~\ref{fig:bestfit}) provides a unique opportunity to investigate an annihilating dark matter interpretation of the observed positron excess without modelling the primary electron spectrum~\cite{Grasso:2009ma} and avoiding the introduction of systematic errors from using $(e^+ + e^-)$ data from the same (or other) experiments~\cite{Kopp:2013eka, Cholis:2013psa, Boudaud:2014dta}.

 In this Appendix, we implement the latest propagation profiles and an updated model for the galactic magnetic field model in the {\fontfamily{cmss}\selectfont{Dragon}}~\cite{dragon:xx} code to solve the propagation equation numerically, obtaining an improved fit for our calculated positron flux in several leptophilic DM annihilation channels.

\subsection{Revisiting the DM interpretation of the positron excess}
\label{sec:DMpositron_a}
In order to interpret the positron spectrum, we first model the sources of (secondary) positrons from astrophysics, then introduce a primary component from DM annihilations to explain
the excess.

\vspace{2mm}
We compute the astrophysical background, or the secondary positron spectrum produced by nuclei in high-energy cosmic rays scattering off of the atoms of the interstellar medium by
using the {\fontfamily{cmss}\selectfont{Dragon}} code \cite{dragon:xx}, for the choice of nuclear CR source parameters described in Appendix \ref{sec:parameters}. We also include an
enhancement factor, $c_{e^+}$, for the secondary positrons to fit the low energy positron flux data. This factor can be viewed as a quantifier of the uncertainties in the effective
proton-proton cross-section due to the presence of heavier nuclei in CRs and in the ISM. Strictly speaking, such a factor would also be a function of energy and spectral index of primary
CRs \cite{Mori:2009te} (see \cite{Kachelriess:2015wpa} for a recent analysis). Our use of a constant factor following \cite{Lin:2014vja}, thus, is an approximation and a more careful
treatment is postponed to a future work.

\vspace{2mm}
The primary component of positron flux is assumed to be comprised of positrons produced from DM annihilations in the Milky Way halo. Due
to a lack of complementary excess, within uncertainties, in the antiproton spectrum \cite{Giesen:2015ufa}, we will consider DM annihilations
in only four channels with leptonic final states: $\mu^ + \mu^- (2\mu)$, $\tau^+ \tau^- (2 \tau)$, $4\mu$ and $4\tau$. The spectra of
positrons per annihilation \textit{at injection} are obtained from the results in {\fontfamily{cmss}\selectfont{PPPC4DMID}}. We note that the
mediators in the {\fontfamily{cmss}\selectfont{PPPC4DMID}} codes for the $4\mu$ and $4\tau$ channels are assumed to be light compared
to the DM mass. The spatial distribution of DM in the galactic halo is assumed to be described by a Navarro-Frenk-White (NFW) density profile \cite{Navarro:1996gj} which is parameterized by a spherically symmetric distribution function,
\beq
\rho (r) = \frac{\rho_s}{(r/r_s)(1 + r/r_s)^2}.
\eeq
Here $r$ is the distance from the galactic center, $\rho_s = 0.26 \textrm{ GeV} \textrm{ cm}^{-3}$ and $r_s = 20 \textrm{ kpc}$ such
that the DM density at Earth is normalized to $\rho_{\odot} = 0.3 \textrm{ GeV} \textrm{ cm}^{-3}$ \cite{Bovy:2012tw}. The DM velocity-
averaged annihilation cross-section ($\left<\sigma v\right>$) and mass ($m_\chi$), which determine the normalization and position of the
peak of the positron flux respectively, are treated as free parameters which we will fit for when comparing the total predicted positron flux to
the data.

\begin{figure}[h!]
                \centering
                \includegraphics[width=\textwidth]{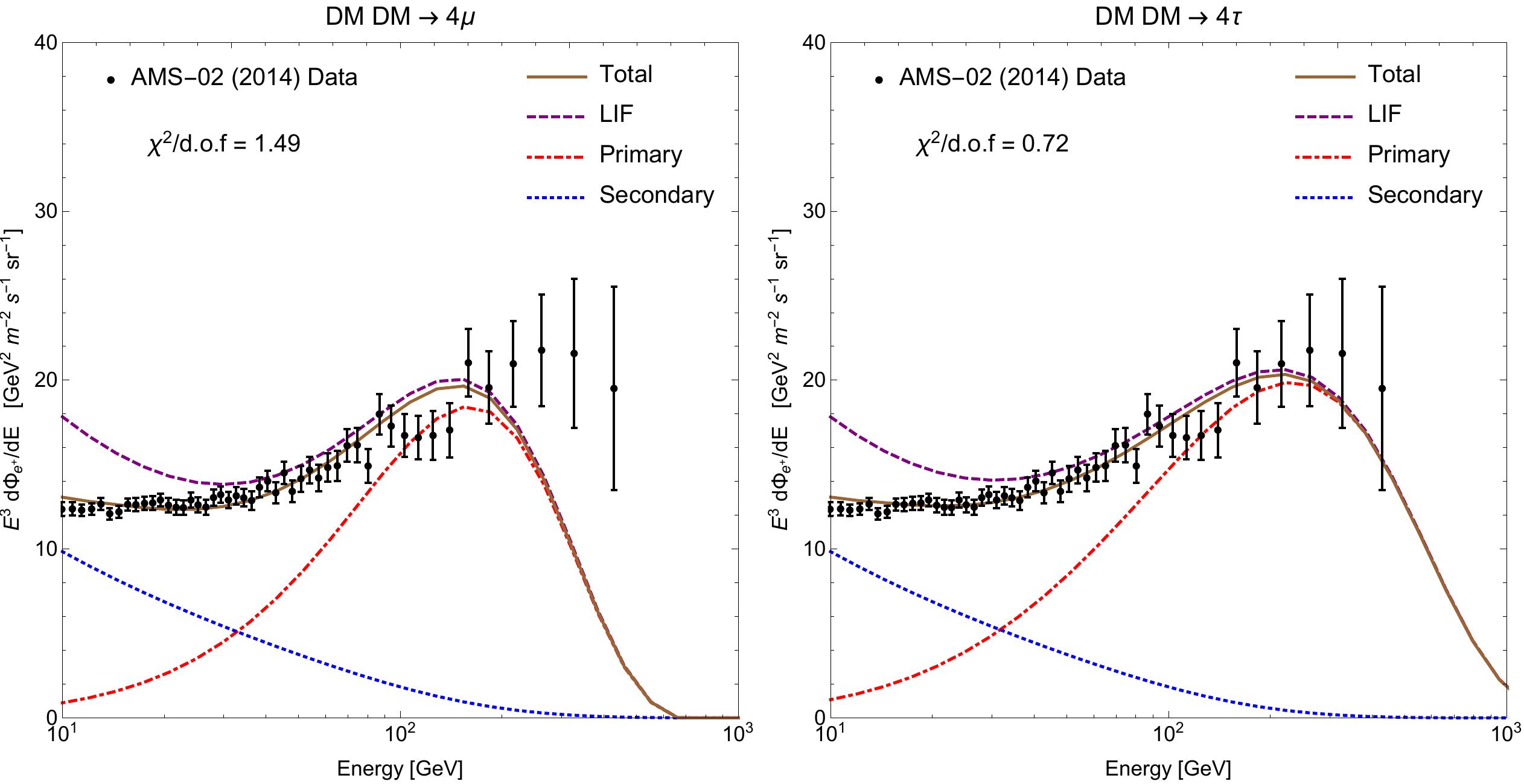}
                \caption{The positron flux as a function of positron energy for the $4\mu$ channel (left) and $4\tau$ channel (right)
                corresponding to the best-fit values of $\left<\sigma v\right>$ and $m_\chi$ for the MODA profile. The $\chi^2_{\text{dof}}$
                for each channel is also indicated in the plot.}
                \label{fig:bestfit}
\end{figure}

\vspace{2mm}
Once we model all sources of positrons at injection, we need to consider the propagation of the positrons from the sources to the earth. Positrons produced in DM annihilations, or from
a secondary astrophysical process, propagate through the turbulent galactic magnetic field (GMF) and get deflected by random inhomogeneities, referred to as Alf\'{v}en waves \cite{Salati:2007zz, Cirelli:2012tf} while also losing energy through several radiative processes. The propagation of positrons can be modelled by a diffusion equation with the
appropriate choices of propagation profiles and GMF models. A propagation profile is defined by a set of parameters that occurs in the diffusion equation
whose values are inferred from fits to data of secondary-to-primary ratios of nuclei in cosmic ray spectra (refer Appendix
\ref{sec:propagation} for a detailed discussion). It is important to note that current precision in the data of secondary-to-primary ratios
cannot help us distinguish between the degeneracies in different propagation profiles. However, the exquisite B/C ratio data presented by
{\fontfamily{cmss}\selectfont {AMS-02}} \cite{ams:2013} indicates that we are entering a new era of precision astrophysics \cite{Genolini:2015cta}, and consequently the current
propagation models (for example: the MIN-MED-MAX profiles \cite{Maurin:2001sj, Maurin:2002hw}
used in \cite{Boudaud:2014dta}) must be updated in accordance with the data. In our analysis, we take a preliminary step in that direction 
by adopting two propagation profiles that incorporate the new {\fontfamily{cmss}\selectfont {AMS-02}} data: MOD A \cite{Lin:2014vja} and
MOD B \cite{Jin:2014ica}, and a `realistic' model for the GMF \cite{Jansson:2012pc}. For all sources (secondaries and dark matter), we implement the propagation in the {\fontfamily{cmss}\selectfont{Dragon}} code, which we use to solve the propagation equation numerically. The total local interstellar positron flux (LIF) after propagation, denoted as $\Phi_{e^+}^{\text{tot,LIF}}(E)$ (indicated by the purple dashed line in Fig.~\ref{fig:bestfit}) is given by a sum
of the DM contribution and the contribution from secondaries,
\beq \label{eq:flux1}
\Phi_{e^+}^{\text{tot, LIF}}(E) = \Phi_{e^+}^{\text{DM}}(E)+ c_{e^+} \Phi_{e^+}^{\text{sec}}(E).
\eeq

In order to obtain the positron flux measured at Earth however, solar modulation effects must also be considered. Various analyses have
focused on developing refined analytical \cite{Potgieter:2013cwj, Cholis:2015gna, DiFelice:2016oec} and numerical \cite{Kappl:2015hxv}
models that incorporate time and charge-dependent effects affecting the low-energy ($\lesssim 10$ GeV) positron spectrum. However, as
solar modulation is a subdominant effect at energies greater than $10$ GeV we work in the force-field approximation \cite{Fisk:1971}, fixing
the value of the Fisk potential to $\phi_\text{F}=600 \, \text{MV}$ \footnote{The value for the Fisk potential quoted in the text is obtained 
by taking the mean of the annual modulation values available at {\url{http://cosmicrays.oulu.fi/phi/Phi_mon.txt}} during the AMS-02 data 
taking period}. Thus, for each annihilation channel for the DM, for a given cross-section and mass, we can compute the solar modulated flux 
of positrons expected at Earth as,
\beq \label{eq:flux2}
\Phi_{e^+}^{\text{tot, modulated}}(E) = \frac{E^2}{(E+\phi_\text{F})^2}\Phi_{e^+}^{\text{tot, LIF}}(E + \phi_\text{F}).
\eeq

\vspace{2mm}
We then proceed to fit the dark matter cross-section and mass by comparing the positron flux computed in \Eq{eq:flux2}, $\Phi_{e^+}^{\text{tot, modulated}}$, to the flux observed by {\fontfamily{cmss}\selectfont {AMS-02}}, $\Phi_{e^+}^{\text{obs}}$. To minimise
the systematic errors from solar modulation effect in our fit, we only include the data above $10$ GeV. We calculate the $\chi^2$ value,
defined as
\beq \label{eq:chi2}
\chi^2 = \sum_i \frac{\left(\Phi_{e^+, i}^{\text{obs}} - \Phi_{e^+, i}^{\text{tot, modulated}}\right)^2}{\sigma_i^2},
\eeq
where the sum in $i$ runs over 49 energy bins of the {\fontfamily{cmss}\selectfont {AMS-02}} data and we take $\sigma_i$ to be the value
of statistical and systematic errors added in quadrature for each bin. Finally, we derive the reduced-$\chi^2$ ($\chi^2_{\text{dof}}$) by
dividing the $\chi^2$ by the degrees of freedom, \textit{i.e.} the number of data-points minus the number of free parameters of the fit.
Initially, we do a crude scan over three parameters $m_\chi$, $\left<\sigma v\right>$ and $c_{e^+}$. We use this to fix the value of
$c_{e^+}$ and then do a more refined scan over the $m_\chi$ and $\left<\sigma v\right>$ values. The $\chi^2$-minimization is performed
separately for the MODA and MODB propagation profiles.

\vspace{2mm}
The best-fit values for $c_{e^+}$, $m_\chi$ and $\left<\sigma v\right>$, based on the minimization of $\chi^2$, have been tabulated in
Tab.~\ref{table:bestfit} for all four leptonic final states that we consider and for both propagation profiles. Using the values in this table, in
Fig.~\ref{fig:bestfit} we plot the primary positron spectrum from DM annihilations in the $4\mu$ and $4\tau$ annihilation channels, the contribution from secondaries, the total LIF calculated in \Eq{eq:flux1}, and the best fit prediction of the total positron flux at Earth after including the solar modulation effects. Fig.~\ref{fig:bestfit} indicates that the low energy bins receive a contribution mostly from the secondary background, while the peaked
primary positron spectra from DM annihilations drives the fit in the high energy bins. Thus, the predicted behavior of a falling positron spectra
above $500 \textrm{ GeV}$ is one of the key features that will help in distinguishing between a DM and pulsar interpretation of the positron
excess \cite{AMS:2015icrc}.

\begin{table}[!h]
\centering
\caption{Best fit values for DM parameters in case of leptophilic annihilation channels. The $\chi^2_{\text{dof}}$ values are computed as
elaborated in the text.} \label{table:bestfit}
\vspace{2mm}
\begin{tabular}{ | E | C | C | A | D | D | }
  \hline			
Propagation Profile & Channel & $m_\chi$ [$\gev$] & $\left<\sigma v\right>$ [$\text{cm}^3/\text{s}$] & $c_{e^+}$ & $\chi^2_{\text{dof}}$\\\hline
  \multirow{4}{*}{MODA} & $\mu^+ \mu^-$ & $340$ & $2.42 \times 10^{-24}$ & $1.80$ & $2.99$ \\
  & $\tau^+ \tau^-$ & $800$ &  $1.73 \times 10^{-23}$ & $1.63$ & $0.68$ \\
  & $Z^\prime Z^\prime \rightarrow 4\mu$ & $570$ & $4.57 \times 10^{-24}$ & $1.72$ &$1.49$ \\
  & $Z^\prime Z^\prime \rightarrow 4\tau$ & $1700$ & $3.63 \times 10^{-23}$ & $1.70$ & $0.72$ \\ \hline
  \multirow{4}{*}{MODB} & $\mu^+ \mu^-$ & $350$ & $2.42 \times 10^{-24}$ & $2.20$ & $3.11$ \\
  & $\tau^+ \tau^-$ & $880$ & $1.77 \times 10^{-23}$ & $2.10$ &$0.76$ \\
  & $Z^\prime Z^\prime \rightarrow 4\mu$ & $570$ & $2.10 \times 10^{-24}$ & $2.10$ &$1.66$ \\
  & $Z^\prime Z^\prime \rightarrow 4\tau$ & $1760$ & $1.70 \times 10^{-23}$ & $2.10$ &$0.80$\\ \hline
\end{tabular}
\end{table}

\vspace{2mm}
Our best fit values for the enhancement factor to the secondary background, $c_{e^+}$, are in close agreement with \cite{Lin:2014vja} for
the MODA profile. The MODB profile provides a good fit to the positron flux data, however the best fit value of $c_{e^+}$ is greater than 2
for all DM annihilation channels, exceeding the maximum expected enhancement for the effective proton-proton cross-section as per
\cite{Mori:2009te}. Thus, we only consider the MODA propagation profile henceforth.

\vspace{2mm}
With reference to the MODA profile, we observe from Tab.~\ref{table:bestfit} that the $2\mu$ annihilation channel is disfavored due to its
poor fit quality, where as the $2\tau$ and $4\tau$ annihilation channels give very good fits to the data due to the softer $e^+$ spectrum
obtained from these final states. We also note that the $4\mu$ channel provides a reasonable fit to the data as well.

\vspace{2mm}
Similar analysis to ours with the latest {\fontfamily{cmss}\selectfont {AMS-02}} data set have been carried out in \cite{Boudaud:2014dta, Lin:2014vja, Jin:2014ica} and our results are in general agreement with theirs. In particular, as we work with the positron flux data with no modelling of the primary electron spectrum, we expect our best fit region in the $m_\chi$ and $\left<\sigma v\right>$ plane to be similar to
the region obtained in \cite{Boudaud:2014dta}\footnote{The authors of \cite{Boudaud:2014dta} first obtain the total predicted positron
flux by adding a primary DM (or pulsar) positron component over the astrophysical background, then use the preliminary $(e^+ + e^-)$
spectra released by {\fontfamily{cmss}\selectfont {AMS-02}} to compute the predicted positron fraction, which they fit to the
{\fontfamily{cmss}\selectfont {AMS-02}} positron fraction data.}, and indeed this is the case. We highlight two differences in our results
however: (1) since we include a complete description of the energy losses and a new, improved model for GMF in the {\fontfamily{cmss}\selectfont{Dragon}} code, this gives us an increased total positron flux, pushing the best fit point towards slightly lower cross-sections, (2) there is an overall improvement in the quality of our fits.

\section{Cosmic Ray Propagation in the Galaxy}
\subsection{Propagation equation and parameters}\label{sec:propagation}
Irrespective of the source of their production, charged cosmic rays (CRs), positrons in our case, propagate through the turbulent magnetic
field in our galaxy where they get deflected by random inhomogeneities. As they propagate they also lose energy through several processes
such as synchrotron radiation, bremsstrahlung, possible reacceleration or convection and from undergoing inverse Compton scattering (ICS)
on the CMB and ambient starlight. Ignoring the losses through fragmentation and radioactive decay, CR transport can be modelled by a
diffusion equation for each CR species \cite{Lin:2014vja},
\beq \label{eq:diffusion}
\frac{\partial \psi}{\partial t} = Q(x, p) + \nabla . (D_{xx} \nabla \psi - V_c \, \psi) + \frac{\partial}{\partial p} p^2 D_{pp}
\frac{\partial}{\partial p} \frac{1}{p^2} \psi - \frac{\partial}{\partial p} \left[\dot{p} \psi - \frac{p}{3}(\nabla. V_c) \psi \right],
\eeq
where $\psi(x,p,t)$ is the spatial density of CRs per unit momentum interval $(p)$ , $Q(x, p)$ is the source distribution, $D_{xx}$ and
$D_{pp}$ are the diffusion coefficients in position and momentum space respectively, $V_c(z)$ is the convection velocity (assumed to vary
only along the height of the galactic disk) and $\dot{p}$ denotes the energy loss coefficient. The spatial diffusion coefficient is
conventionally taken to be position-independent and is parameterized by a power law $D_{xx} = D_0 \, (R/R_0)^\delta$, where $R \equiv pc/Ze$ is the rigidity of the particle and $\delta$ is the power law dependence of $D$ on energy. The coefficient of diffusive reacceleration
in momentum space, $D_{pp}$, is a function of the velocity of hydrodynamical disturbances also called the Alfv\`{e}n wave speed, $v_A$.
The energy loss coefficient, $\dot{p}$, takes into account all energy losses mentioned above (for details of the various contributions see
Ref.~\cite{Buch:2015iya}).

\vspace{2mm}
\Eq{eq:diffusion} can be solved numerically using codes such as {\fontfamily{cmss}\selectfont{Galprop}} \cite{galprop:yy} and
{\fontfamily{cmss}\selectfont{Dragon}} \cite{dragon:xx}, or semi-analytically which gives a better handle on the input parameters. Both
these techniques proceed by modelling the diffusive region of the galaxy as a two-dimensional cylinder extending radially upto $20 \textrm{ kpc}$, with height $2 L$ in the vertical direction. The boundary condition is imposed such that the quantity $\psi$ vanishes at $z=\pm L$
implying that CRs can escape freely beyond the boundaries of the cylinder. The height of the cylinder $L$ can vary between $1 \hyphen 15 \textrm{ kpc}$ for different propagation models, however, recent combined analyses of synchrotron emission and $\gamma$-ray data
strongly disfavour models with $L<2 \textrm{ kpc}$ \cite{DiBernardo:2012zu, Orlando:2013ysa}.

\vspace{2mm}
Notice that we still need to fix the major propagation parameters $(D_0, \delta, L, dV_c/dz, v_A)$. This is done by fitting-to-data the
secondary-to-primary Boron-to-Carbon (B/C) ratio and the unstable-to-stable Beryllium ratio ($^{10}\text{Be}/^{9}\text{Be}$) in cosmic rays
as they are largely independent of the injection spectra, and then extracting the best fit values of the parameters for each propagation model. While
computing the value of these parameters by doing a global Bayesian fit using statistical techniques such as the Markov Chain Monte-Carlo (MCMC)
scan \cite{Putze:2010zn, Trotta:2010mx} minimizes the systematic error in an analysis, popular models available in the market, the MIN-MED-MAX
profiles \cite{Maurin:2001sj, Maurin:2002hw} for instance, are generally adopted for convenience or for consistency with previous results \cite{Buch:2015iya}. In our analysis, we adopt two benchmark models for propagation from \cite{Lin:2014vja} and \cite{Jin:2014ica}. The parameters
appearing in the diffusion equation for the models we consider are presented in Tab.~\ref{table:propagation1}. The models themselves
are briefly described below.
\begin{itemize}
\item Model A (MODA) is a propagation model that also includes \textit{diffusive convection} along with other energy-loss mechanisms. So,
there is an additional parameter for the galactic convective wind, $V_c$, that is switched on in the global fit.

\item Model B (MODB) includes \textit{diffusive reacceleration} instead of convection. Motivated by \cite{Putze:2010zn, Trotta:2010mx},
the authors of \cite{Jin:2014ica} use reacceleration (implemented in {\fontfamily{cmss}\selectfont{Dragon}} through the ``Ptsukin03''
parametrization derived from \cite{Ptuskin:2003zv}) as a mechanism to reproduce the observed peak in the B/C spectrum.
\end{itemize}

\begin{table}[!h]
\centering
\caption{Propagation parameters for profiles MODA, MODB based on \cite{Lin:2014vja} and \cite{Jin:2014ica} respectively.}\label{table:propagation1}
\vspace{2mm}
\begin{tabular}{ | l | B | B | }
  \hline			
Parameters & Model A & Model B \\ \hline
$D_0 \, (10^{28} \text{cm}^2/\text{s})$ & $1.95$ & $6.464 $ \\
$\delta$ & $0/0.51$ & $0.29$ \\
$R_{\text{0}}$~(GV) & 4.71 & 4 \\
$L \, (\textrm{kpc})$ & $2.5$ & $3.2$ \\
$dV_c/dz \, (\text{km}/\text{s})$ & $4.2$& - \\
$v_A  \, (\text{km}/\text{s})$ & - & $44.7$ \\ \hline
\end{tabular}
\end{table}

\subsection{Parameters for Input Injection Spectra} \label{sec:parameters}
The total positron flux can be expressed as the sum of a secondary astrophysical background and a primary component from DM annihilations.
The secondary component is produced by the spallations of CR nuclei on the ISM. In order to compute the secondary positron component, we
first need to set the nuclear CR source parameters in {\fontfamily{cmss}\selectfont{Dragon}} as described below. The steady state CR flux
after propagation is then used to numerically compute the secondary positron source term $(Q_{e^{+}})$ by following the parametrization
for secondary pion production in spallations developed by \cite{Kamae:2006bf}.

\vspace{2mm}
The source term for every CR nuclear species is given by
\beq \label{eq:source}
Q_i (E_k, r, z) = f_{S}(r, z) \, N_{i} \left( \frac{R}{R_{\text{br}}} \right)^{-\gamma_{i}},
\eeq
where $i$ denotes the nuclear species, $R$ is the rigidity of the particle, $\gamma$ is a power law index that may have different values
above and below $R_{\text{br}}$, and $N_i$ is the normalization fixed according to the observed CR data. Assuming the injected CR
particles are produced in supernova remnants (SNR), the spatial distribution factor $(f_{S})$ should follow the SNR distribution and is given in
cylindrical co-ordinates by,
\beq
f_{S}(r, z) = \left( \frac{r}{r_\odot} \right)^{a} \exp \left( -b.\frac{r - r_\odot}{r_\odot} - \frac{|z|}{z_s} \right).
\eeq
Here $r_{\odot} = 8.5 \, \kpc$ is the Sun's distance from the galactic center, and $z_s = 0.2 \, \kpc$ is the height of the galactic disk.
$a=1.25$ and $b=3.56$ are fixed to reproduce the distribution of galactic CR sources based on {\fontfamily{cmss}\selectfont {Fermi-LAT}}
$\gamma$-ray data, following the prescription in \cite{Trotta:2010mx}. In our analysis the nuclear injection parameters in \Eq{eq:source} are
set to their best-fit values derived by fitting to the latest {\fontfamily{cmss}\selectfont {AMS-02}} proton data \cite{Aguilar:2015ooa}, and
these values are shown in Tab.~\ref{table:propagation2} for the MODA and MODB propagation profiles.

\begin{table}[!h]
\centering
\caption{Source parameters for profiles MODA, MODB based on \cite{Lin:2014vja} and \cite{Jin:2014ica} respectively.} \label{table:propagation2}
\vspace{2mm}
\begin{tabular}{ | l | B | B | }
  \hline			
Parameters & Model A & Model B \\ \hline
$R_{\text{br}}~(GV)$ & $10$ & $12.88$ \\
$\gamma_p$ & $2.336$ & $1.79/2.45$\\
$N_p \, (\text{m}^{-2} \text{s}^{-1} \text{sr}^{-1} \gev^{-1})$ & $0.04783$ & $0.0483$ \\ \hline
\end{tabular}
\end{table}

\newpage
\bibliographystyle{utphys}
\bibliography{dmindirect}

\end{document}